\documentclass[preprint,aps,prd,showpacs,nofootinbib]{revtex4}
\usepackage{mathrsfs}
\usepackage{amsmath}
\usepackage{graphicx}
\usepackage{subfigure}
\usepackage{bm}
\usepackage{epstopdf}
\usepackage{float}

\newcommand{\bea}{\begin{eqnarray}}
\newcommand{\eea}{\end{eqnarray}}
\newcommand{\beq}{\begin{equation}}
\newcommand{\eeq}{\end{equation}}
\newcommand{\nn}{\nonumber}

\def\/{\over}

\begin{document}
\title{\bf Energy shift and Casimir-Polder force for an atom out of thermal equilibrium near a dielectric substrate}
\author{Wenting Zhou and Hongwei Yu }
\affiliation{$^1$ Center for Nonlinear Science and Department of Physics, Ningbo
University, Ningbo, Zhejiang 315211, China
}

\begin{abstract}
We study the energy shift and the Casimir-Polder force of an atom out of thermal equilibrium near the surface of a dielectric substrate.
We first generalize, adopting the local source hypothesis,  the formalism proposed by Dalibard, Dupont-Roc and Cohen-Tannoudji, which separates the contributions of thermal fluctuations and radiation reaction to the energy shift and allows a distinct treatment to atoms in the ground and excited states, to the case out of thermal equilibrium, and then we use the generalized formalism to calculate the energy shift and the Casimir-Polder force of an isotropically polarizable neutral  atom.  We identify the effects of the thermal fluctuations that originate from the substrate and the environment and  discuss in detail how the Casimir-Polder force out of thermal equilibrium behaves in three different distance regions in both the low-temperature limit and the high-temperature limit for both the ground-state and excited-state atoms, with special attention devoted to the new features as opposed to  thermal equilibrium.  In particular, we recover the new behavior of the atom-wall force out of thermal equilibrium at large distances in the low temperature limit recently found in a different theoretical framework and furthermore we give a concrete region where this behavior holds.

\end{abstract}
\pacs{31.30.jh, 12.20.-m, 34.35.+a, 42.50.Nn}
\maketitle

\section{Introduction}

The effect of interaction between an atom and quantum electromagnetic fields has been a long-standing subject of research. It is well-known that even in vacuum, the energy levels of an atom are slightly shifted as  a result of the interaction between the atom and the  fluctuating vacuum  electromagnetic fields~\cite{Lamb47}, and these shifts are further modified  when boundaries which confine the fields  appear. In fact,   when the fluctuations of quantum fields are altered by the presence of boundaries, many novel effects may arise, such as the Casimir effect \cite{C-P48}, the light-cone
fluctuations when gravity is quantized \cite{YF99,YF00,YW,YFS09}, and
the Brownian (random) motion of test particles in an
electromagnetic vacuum \cite{YF04,YC,YCW06,TY} (also see
\cite{JR,Barton0,JR1}), just to name a few.

In 1948, Casimir and Polder discovered that a neutral atom near a perfectly conducting wall feels a net force as a result of the interaction between the atom and vacuum electromagnetic fluctuations~\cite{C-P48}. At short distances, the force behaves like the van der Waals-London interatomic force which decays as $1/z^4$ where $z$ is the distance from the wall, while at large distances, the inclusion of  relativistic retardation effects yields a different $1/z^5$ dependence and this region is called the Casimir-Polder regime. Subsequently, by employing the theory of electromagnetic fluctuations developed by Rytov~\cite{Rytov}, Lifshitz showed that besides the zero-point fluctuations, the thermal fluctuations also give rise to  a revision to the atom-wall force~\cite{Lifshitz56,Lifshitz61}  which actually becomes the leading contribution to the total force at distances much larger than the wavelength of thermal photons and  decays as $T/z^4$. Later, it was shown that the thermal fluctuations also alter the energy shifts of an atom~\cite{Barton}. In recent years,  the research on the Casimir-Polder force has been extended to various circumstances, such as in the presence of partly or perfectly reflecting boundaries in the vicinity of an atom which is static or in non-inertial motion in vacuum~\cite{Passante98,Rizzutu07,Messina,Rizzutu09,Z-Y10} or immersed in a thermal bath~\cite{Eberlein,Z-Y09,S-Y10,Z-Y12}.

The effect of the thermal fluctuations on the Casimir-Polder force referred to above is  about an atom-wall system in thermal equilibrium. Recently, there has been growing interest in the Casimir-Polder force of an atom out of thermal equilibrium  both theoretically and experimentally~\cite{Henkel,Antezza05,Antezza06,Obrecht,Antezza08,Ellingsen09,Buhmann08,Buhmann091,Buhmann092,Sherkunov09}. In Refs.~\cite{Antezza05,Antezza06},  Antezza etal  calculate,  using the fluctuational electrodynamics developed by Rytov et. al~\cite{Rytov} and the linear response theory by Wiley and Sipe~\cite{Wylie}, the Casimir-Polder force felt by an atom near the surface of a half-space dielectric substrate whose temperature is different from that of the thermal bath in the other half-space (environment) under the assumption that the whole system is a stationary configuration, and they find that the force exhibits a new behavior at very large distances when the temperature is low, which decays more slowly with the distance than at the thermal equilibrium. The force is  also distinctive from that in the case of thermal equilibrium as it displays a sizable temperature dependence which could be attractive or repulsive depending on whether the temperature of the substrate is higher or lower than that of the environment. It is interesting to note that this new behavior has already been demonstrated in experiment~\cite{Obrecht}.

In this paper, we study the energy shift and the Casimir-Polder force of an atom near a  dielectric substrate out of thermal equilibrium using a QED treatment of the atom-field coupling.  In such a framework,  on the one hand,  the fluctuating field which is modified by the appearance of the substrate disturbs the atom, and on the other hand, the disturbed atom induces a radiative field in reaction to the disturbance, and both these fields affect  the dynamics of the atom. It has been found in QED that to what extent each mechanism plays a part is determined by the ordering between the operators of the atom and the field in the interaction Hamiltonian~\cite{Ackerhalt73,Senitzky,Milonni73,Milonni75}.  In other words, there exists an indetermination in the contribution of vacuum fluctuations and radiation reaction. The ambiguity was resolved when  Dalibard, Dupont-Roc and Cohen-Tannoudji (DDC) showed that there exists a preferred symmetric operator ordering which enssures that the distinct contributions of vacuum fluctuations and the radiation reaction of the atom to the rate of change of the atomic observables are separately Hermitian~\cite{DDC82,DDC84}. Recently, this formalism has been employed to study the radiative properties of atoms in various cases including non inertial  motion of the atom and a thermal bath at equilibrium~\cite{Audretsch94,Audretsch95, Yu-Lu05, Yu-Zhu06,Zhu-Yu06,Passante98,Rizzutu07,Messina,Rizzutu09,Z-Y10, Z-Y09,S-Y10}.  In the present paper, we will first generalize the DDC formalism originally established  for  thermal equilibrium to the
case out of thermal equilibrium in a stationary regime and then perform a systematic analysis of the atom-wall force for an atom near a dielectric substrate which was examined before by other authors only for atoms in the ground state  in the low temperature limit at very large distances~\cite{Antezza05,Antezza06}. The DDC formalism based  upon the atom-field coupling separates the contributions of thermal fluctuations (including vacuum fluctuations) and the radiation reaction and allows a distinct microscopic treatment to atoms in the ground and excited states, and it differs from  the macroscopic approach  using Lifshitz theory where atoms are treated as a limiting case of a dielectric~\cite{Antezza06-1,Antezza08} and the linear response description of the atom~\cite{Antezza05,Antezza06}. With the DDC formalism to be generalized to the atom-wall system out of thermal equilibrium,  we are able to derive the Casimir-Polder force for an atom out of thermal equilibrium at all distance regimes in both the high- and low-temperature limits for both the ground  and excited states. In particular, we quantify the region of ``very large distances"  which was taken as mathematical  infinity ($z \rightarrow \infty$) in~\cite{Antezza05,Antezza06}, where the new behavior of the force with a weaker distance dependence characterized by $1/z^3$ appears. In other words, we give a concrete region where this new behavior holds.

The paper is organized as follows.  In section II, we briefly review the quantum electromagnetic fields concerned with a general dielectric substrate. In section III, we generalize the DDC formalism to the case  out of thermal-equilibrium. In section IV,  we calculate the energy shift of a two-level atom  near a  dielectric substrate, separating  the contributions of the thermal fluctuations and radiation reaction using the generalized DDC formalism. In section V, we discuss the atomic energy shift and the Casimir-Polder force  near a non-dispersive real dielectric substrate,  and  we summarize in section VI.

\section{The quantum electromagnetic field}

In the presence of magnetoelectric background media where no external sources and currents appear, the classical electromagnetic fields satisfy the Maxwell equations
\bea
\left\{
  \begin{array}{ll}
    \bigtriangledown\cdot\mathbf{B}(t,\mathbf{r})=0, \\
    \bigtriangledown\times\mathbf{E}(t,\mathbf{r})=-{{\partial\mathbf{B}(t,\mathbf{r})}\/{\partial t}}, \\
    \bigtriangledown\cdot\mathbf{D}(t,\mathbf{r})=0, \\
    \bigtriangledown\times\mathbf{H}(t,\mathbf{r})={{\partial\mathbf{D}(t,\mathbf{r})}\/{\partial t}}\;.
  \end{array}
\label{Maxwell equations}
\right.
\eea
By performing the Fourier transformation which is defined for an arbitrary operator, $O(t,\mathbf{r})$,
as
\beq
O(t,\mathbf{r})=\int^{\infty}_{-\infty}d\omega\;e^{-i\omega t}O(\mathbf{r},\omega)\;,\label{fourier transformation}
\eeq
the Maxwell equations can be transformed to
\bea
\left\{
  \begin{array}{ll}
    \bigtriangledown\cdot\mathbf{B}(\mathbf{r},\omega)=0, \\
    \bigtriangledown\times\mathbf{E}(\mathbf{r},\omega)=i\omega\mathbf{B}(\mathbf{r},\omega), \\
    \bigtriangledown\cdot\mathbf{D}(\mathbf{r},\omega)=0, \\
    \bigtriangledown\times\mathbf{H}(\mathbf{r},\omega)=-i\omega\mathbf{D}(\mathbf{r},\omega)\;.
  \end{array}
\label{transformed Maxwell equations}
\right.
\eea
Assuming that the medium under consideration is not bi-anisotropic, we can express the electric displacement vector $\mathbf{D}(\mathbf{r},\omega)$ and  the magnetic field strength $\mathbf{H}(\mathbf{r},\omega)$  as
\bea
\mathbf{D}(t,\mathbf{r})&=&\varepsilon_0\mathbf{E}(t,\mathbf{r})+\mathbf{P}(t,\mathbf{r}), \label{D}\\
\mathbf{H}(t,\mathbf{r})&=&{\mathbf{B}(t,\mathbf{r})\/\mu_0}-\mathbf{M}(t,\mathbf{r})\label{H}
\eea
where $\varepsilon_0$ and $\mu_0$ are the permittivity and permeability of vacuum, and $\mathbf{P}(t,\mathbf{r})$ and $\mathbf{M}(t,\mathbf{r})$ are the polarization and magnetization fields respectively. Particularly, for the medium which responds linearly and locally to externally applied fields, the most general relations between the fields that are consistent with causality and the linear fluctuation-dissipation theorem can be written as
\bea
\mathbf{P}(t,\mathbf{r})&=&\varepsilon_0\int^{\infty}_0d\tau\chi_e(\tau,\mathbf{r})\mathbf{E}(t-\tau,\mathbf{r})
+\mathbf{P}_N(t,\mathbf{r}), \label{part D}\\
\mathbf{M}(t,\mathbf{r})&=&{1\/\mu_0}\int^{\infty}_0d\tau\chi_m(\tau,\mathbf{r})\mathbf{B}(t-\tau,\mathbf{r})
+\mathbf{M}_N(t,\mathbf{r}) \label{part H}
\eea
where $\mathbf{P}_N(t,\mathbf{r})$ and $\mathbf{M}_N(t,\mathbf{r})$ are respectively the noise polarization and magnetization associated with the absorption of the medium with electric and magnetic susceptibilities $\chi_e(\tau,\mathbf{r})$ and $\chi_m(\tau,\mathbf{r})$. Plugging the above two equations into Eqs.~(\ref{D}) and (\ref{H}), and then performing the Fourier transform (\ref{fourier transformation}) for the operators concerned, we obtain
\bea
\mathbf{D}(\mathbf{r},\omega)&=&\varepsilon_0\epsilon(\mathbf{r},\omega)\mathbf{E}(\mathbf{r},\omega)
+\mathbf{P}_N(\mathbf{r},\omega)\;, \label{D transformed}\\
\mathbf{H}(\mathbf{r},\omega)&=&\kappa_0\kappa(\mathbf{r},\omega)\mathbf{B}(\mathbf{r},\omega)
-\mathbf{M}_N(\mathbf{r},\omega)
\label{H transformed}
\eea
with $\kappa_0={\mu_0}^{-1}$ and
\bea
\epsilon(\mathbf{r},\omega)&=&1+\int^{\infty}_0d\tau\;\chi_e(\tau,\mathbf{r})\;e^{i\omega\tau}\;,\\
\kappa(\mathbf{r},\omega)&=&1-\int^{\infty}_0d\tau\;\chi_m(\tau,\mathbf{r})\;e^{i\omega\tau}\;,
\eea
which are called the relative permittivity and permeability respectively. The insertion of Eqs.~(\ref{D transformed}) and (\ref{H transformed}) into Equations (\ref{transformed Maxwell equations}) yields
\bea
\left\{
  \begin{array}{ll}
    \bigtriangledown\cdot\mathbf{B}(\mathbf{r},\omega)=0, \\
    \bigtriangledown\times\mathbf{E}(\mathbf{r},\omega)=i\omega\mathbf{B}(\mathbf{r},\omega), \\
    \varepsilon_0\bigtriangledown\cdot[\epsilon(\mathbf{r},\omega)\mathbf{E}(\mathbf{r},\omega)]=
\rho_N(\mathbf{r},\omega), \\
    \bigtriangledown\times[\kappa(\mathbf{r},\omega)\mathbf{B}(\mathbf{r},\omega)]
+i{\omega\/c^2}\epsilon(\mathbf{r},\omega)\mathbf{E}(\mathbf{r},\omega)=\mu_0\mathbf{j}_N(\mathbf{r},\omega)\;,
  \end{array}
\label{transformed Maxwell equations 1}
\right.
\eea
where
\bea
\rho_N(\mathbf{r},\omega)&=&-\bigtriangledown\cdot\mathbf{P}_N(\mathbf{r},\omega)\;,\\
\mathbf{j}_N(\mathbf{r},\omega)&=&-i\omega\mathbf{P}_N(\mathbf{r},\omega)
+\bigtriangledown\times\mathbf{M}_N(\mathbf{r},\omega)\;.
\eea
For a non-magnetic medium, $\kappa(\mathbf{r},\omega)=1$, $\mathbf{j}_N(\mathbf{r},\omega)=-i\omega\mathbf{P}_N(\mathbf{r},\omega)$. Combining these relations with the second and the fourth equations in Equations (\ref{transformed Maxwell equations 1}), we arrive at the differential equation satisfied by the electric field,
\beq
\bigtriangledown\times\bigtriangledown\times\mathbf{E}(\mathbf{r},\omega)
-{\omega^2\/c^2}\epsilon(\mathbf{r},\omega)\mathbf{E}(\mathbf{r},\omega)=
i\mu_0\omega\mathbf{j}_N(\mathbf{r},\omega)\;.\label{differential equation of E}
\eeq
The solution of this equation can be expressed in terms of the dyadic Green's function,
$\texttt{G}(\mathbf{r},\mathbf{r}',\omega)$, as
\beq
\mathbf{E}(\mathbf{r},\omega)=i\mu_0\omega\int d^3\mathbf{r}'\;\texttt{G}(\mathbf{r},\mathbf{r}',\omega)\cdot\mathbf{j}_N(\mathbf{r}',\omega)\;.
\label{E in terms of the G function}
\eeq
 Substitution of the above equation into Eq.~(\ref{differential equation of E}) leads to the differential equation for the Green's function
\beq
\{\partial^r_i\partial^r_m-\delta_{im}[\bigtriangleup^r
+\epsilon(\mathbf{r},\omega){\omega^2\/c^2}]\}G_{mj}(\mathbf{r},\mathbf{r}',\omega)
=\delta_{ij}\delta(\mathbf{r}-\mathbf{r}')
\eeq
where $\bigtriangleup^r=\partial^r_i\partial^r_i$. Hereafter, the Einstein summation convention is assumed for repeated indices.

So far, all the discussions regard the classical electrodynamics. However, we need  a theory of quantized electromagnetic fields in a dielectric medium for our purpose.  In this regard,  let us note that the quantization of the electromagnetic field in an absorbing dielectric has been widely discussed~\cite{Matloob95,Gruner96,Dung}. In this paper, we are concerned with a type of nonmagnetic medium with
\beq
\epsilon(\mathbf{r},\omega)=\epsilon_R(\mathbf{r},\omega)+i\epsilon_I(\mathbf{r},\omega)\;,\quad \kappa(\mathbf{r},\omega)=1\;.
\eeq
Following Refs.~\cite{Matloob95,Gruner96}, $\mathbf{j}_N(\mathbf{r},\omega)$ can be related to a bosonic vector field, $\mathbf{a}(\mathbf{r},\omega)$, as
\beq
\mathbf{j}_N(\mathbf{r},\omega)={\omega\/{\mu_0c^2}}
\sqrt{{\hbar\/{\pi\varepsilon_0}}\epsilon_I(\mathbf{r},\omega)}\;\mathbf{a}(\mathbf{r},\omega)
\label{Quantified j}
\eeq
with the vector operator $\mathbf{a}(\mathbf{r},\omega)$ and its Hermitian conjugates satisfying the following commutation relations
\bea
&&[a_i(\mathbf{r},\omega),a^{\dag}_j(\mathbf{r}',\omega')]=
\delta_{ij}\delta(\mathbf{r}-\mathbf{r}')\delta(\omega-\omega')\;,\\
&&[a_i(\mathbf{r},\omega),a_j(\mathbf{r}',\omega')]=0=
[a^{\dag}_i(\mathbf{r},\omega),a^{\dag}_j(\mathbf{r}',\omega')]\;.
\eea
Putting Eq.~(\ref{Quantified j}) into Eq.~(\ref{E in terms of the G function}), the field operator can be re-expressed as
\beq
\mathbf{E}(\mathbf{r},\omega)=i\sqrt{{\hbar\/{\pi\varepsilon_0}}}{\omega^2\/c^2}\int d^3\mathbf{r}'\sqrt{\epsilon_I(\mathbf{r}',\omega)}\;\texttt{G}(\mathbf{r},\mathbf{r}',\omega)
\cdot\mathbf{a}(\mathbf{r}',\omega)\;.\label{E-G-f}
\eeq
It is explicit that the spatial distribution of the electric field is determined by the dyadic Green's function, $\texttt{G}(\mathbf{r},\mathbf{r}',\omega)$, which is determined by the spatial distribution of the medium.

For a  configuration with one half-space ($z<0$) occupied by a dielectric substrate and the other half-space ($z>0$) being an empty space, which is of  particular interest in the present paper, the components of the dyadic Green's function are~\cite{Dung}
\beq
\mathrm{G}_{ij}(\mathbf{r},\mathbf{r}',\omega)=
\left\{
     \begin{array}{ll}
     \mathrm{G}^0_{ij}(\mathbf{r},\mathbf{r}',\omega)+\mathrm{R}_{ij}(\mathbf{r},\mathbf{r}',\omega), \quad\;z>0,\;z'>0\;, \\
     \mathrm{T}_{ij}(\mathbf{k}_{\parallel},\omega,z,z'), \quad\;\quad\;\quad\;\quad\quad z>0,\;z'<0\;.
     \end{array}
\right.
\label{green function}
\eeq
where $\mathrm{G}^0_{ij}(\mathbf{r},\mathbf{r}',\omega)$ corresponds to the Green's function of a vacuum that is Fourier transformed, $\mathrm{R}_{ij}(\mathbf{r},\omega)$ and $\mathrm{T}_{ij}(\mathbf{r},\omega)$ describe the reflection and transmission at the interface and they can be expanded as follows
\bea
\mathrm{R}_{ij}(\mathbf{r},\mathbf{r}',\omega)
&=&\int{d^2\mathbf{k}_{\parallel}\/4\pi^2}\mathrm{R}_{ij}(\mathbf{k}_{\parallel},\omega,z,z')
e^{i\mathbf{k}_{\parallel}\cdot(\mathbf{r}_{\parallel}-\mathbf{r}'_{\parallel})}\;,\\
\label{R integration}
\mathrm{T}_{ij}(\mathbf{r},\mathbf{r}',\omega)
&=&\int{d^2\mathbf{k}_{\parallel}\/4\pi^2}\mathrm{T}_{ij}(\mathbf{k}_{\parallel},\omega,z,z')
e^{i\mathbf{k}_{\parallel}\cdot(\mathbf{r}_{\parallel}-\mathbf{r}'_{\parallel})}\;,
\label{R integration}
\eea
where $\mathbf{k}_{\parallel}=(k_x,k_y,0)$, $\mathbf{r}_{\parallel}=(x,y,0)$ are two-dimensional
vectors in the $(x,y)$ plane,
\bea
\left\{
  \begin{array}{ll}
  \mathrm{R}_{xx}={i\/{2\beta_2}}e^{i\beta_2(z+z')}
  [{r^p_{21}\/q_2^2}(-\beta_2^2{k_x^2\/k_{\parallel}^2})+r^s_{21}{k_y^2\/k_{\parallel}^2}]\;,\\
  \mathrm{R}_{xy}={i\/{2\beta_2}}e^{i\beta_2(z+z')}
  [{r^p_{21}\/q_2^2}(-\beta_2^2{k_x k_y\/k_{\parallel}^2})-r^s_{21}{k_x k_y\/k_{\parallel}^2}]\;,\\
  \mathrm{R}_{xz}={i\/{2\beta_2}}e^{i\beta_2(z+z')}{r^p_{21}\/q_2^2}
  [-\beta_2k_x]\;,\\
  \mathrm{R}_{yx}=\mathrm{R}_{xy}\;,\\
  \mathrm{R}_{yy}=\mathrm{R}_{xx}(k_x\leftrightarrow k_y)\;,\\
  \mathrm{R}_{yz}=\mathrm{R}_{xz}(k_x\leftrightarrow k_y)\;,\\
  \mathrm{R}_{zx}=-\mathrm{R}_{xz}\;,\\
  \mathrm{R}_{zy}=-\mathrm{R}_{yz}\;,\\
  \mathrm{R}_{zz}={i\/{2\beta_2}}e^{i\beta_2(z+z')}{r^p_{21}\/q_2^2}k_{\parallel}^2\;,
  \end{array}
\right.
\eea
and
\bea
\left\{
  \begin{array}{ll}
  \mathrm{T}_{xx}={i\/{2\beta_2}}e^{i(\beta_2z-\beta_1z_1)}
  ({t^p_{21}\/q_2q_1}\beta_2\beta_1{k_x^2\/k_{\parallel}^2}+t^s_{21}{k_y^2\/k_{\parallel}^2})\;,\\
  \mathrm{T}_{xy}={i\/{2\beta_2}}e^{i(\beta_2z-\beta_1z_1)}
  ({t^p_{21}\/q_2q_1}\beta_2\beta_1{k_x k_y\/k_{\parallel}^2}-t^s_{21}{k_x k_y\/k_{\parallel}^2})\;,\\
  \mathrm{T}_{xz}={i\/{2\beta_2}}e^{i(\beta_2z-\beta_1z_1)}{t^p_{21}\/q_2q_1}(-\beta_2k_x)\;,\\
  \mathrm{T}_{yx}=\mathrm{T}_{xy}\;,\\
  \mathrm{T}_{yy}=\mathrm{T}_{xx}(k_x\leftrightarrow k_y)\;,\\
  \mathrm{T}_{yz}=\mathrm{T}_{xz}(k_x\leftrightarrow k_y)\;,\\
  \mathrm{T}_{zx}={i\/{2\beta_2}}e^{i(\beta_2z-\beta_1z_1)}{t^p_{21}\/q_2q_1}(-\beta_1k_x)\;,\\
  \mathrm{T}_{zy}=\mathrm{T}_{zx}(k_x\leftrightarrow k_y)\;,\\
  \mathrm{T}_{zz}={i\/{2\beta_2}}e^{i(\beta_2z-\beta_1z_1)}{t^p_{21}\/q_2q_1}k_{\parallel}^2
  \end{array}
\right.
\eea
with
\bea
&&q_1(\omega)={\omega\/c}\sqrt{\epsilon}\;,\quad\quad q_2(\omega)={\omega\/c}\;,\\
&&\beta_{\alpha}\equiv\beta_{\alpha}(\omega)=\sqrt{q_\alpha^2(\omega)-k_{\parallel}^2}\;,\quad\;
(\alpha=(1,2),\;\mathrm{Re}[\beta_{\alpha}]\geq 0,\; \mathrm{Im}[\beta_{\alpha}]\geq 0)\;,\\
&&r^p_{21}={{\epsilon\beta_2-\beta_1}\/{\epsilon\beta_2+\beta_1}}\;,\quad\;
r^s_{21}={{\beta_2-\beta_1}\/{\beta_2+\beta_1}}\;,\\
&&t^p_{21}={{2\sqrt{\epsilon}\beta_2}\/{\epsilon\beta_2+\beta_1}}\;,\quad\;
t^s_{21}={{2\beta_2}\/{\beta_2+\beta_1}}\;.
\label{parameters}
\eea
Here we have denoted $\mathrm{R}_{ij}(\mathbf{k}_{\parallel},\omega,z,z')$ and $\mathrm{T}_{ij}(\mathbf{k}_{\parallel},\omega,z,z')$ by $\mathrm{R}_{ij}$ and $\mathrm{T}_{ij}$ for simplicity.

In the following, we calculate the energy shift and the Casimir-Polder force of an atom  near a half-space dielectric substrate which is locally at thermal equilibrium at a temperature that is different from the temperature of the environment (empty space with thermal radiation) where the atom is located. To do so,  we should first generalize the DDC formalism to the case out of thermal-equilibrium.

\section{The generalized DDC formalism}

Consider an atom in interaction with quantum electromagnetic fields. Let $\tau$ denote the atomic proper time and $x(\tau)=(t(\tau),\textbf{r}(\tau))$ represent the  stationary atomic  trajectory. The stationarity of the trajectory guarantees the existence of stationary states of the atom.
The Hamiltonian that governs the evolution of the atom is
\beq
H_A(\tau)=\hbar\sum_n\omega_n\sigma_{nn}(\tau)
\eeq
where $\sigma_{nn}=|n\rangle\langle n|$. The Hamiltonian of the free electromagnetic field with respect to $\tau$ is
\beq
H_F(\tau)=\int d^3\mathbf{r}\int^{\infty}_0d\omega\;\hbar\omega\;
a^{\dag}_{i}(t,\mathbf{r},\omega)a_{i}(t,\mathbf{r},\omega){dt\/d\tau}\;.
\eeq
In the multipolar coupling scheme~\cite{Dung,Compagno95}, the Hamiltonian that describes the interaction between the atom and the field is given by
\beq
H_I(\tau)=-\bm{\mu}(\tau)\cdot\mathbf{E}(x(\tau))=
-\sum_{mn}\bm{\mu}_{mn}\cdot\mathbf{E}(x(\tau))\sigma_{mn}(\tau)
\eeq
where $\bm{\mu}$ is the electric dipole moment of the atom. The total Hamiltonian of the system (atom + field) is composed of the above three parts
\beq
H(\tau)=H_A(\tau)+H_F(\tau)+H_I(\tau)\;.
\eeq
Starting from the above Hamiltonian, we can write out the Heisenberg equations of motion for the dynamical variables of the atom and the field, and up to the first order of the coupling constant $\mu$, the solutions of each equation can then be divided into two parts: a free part that exists even when there is no coupling between the atom and the field and corresponds to the effect of the thermal fluctuations (including vacuum fluctuations), and a source part that is induced by the interaction between the atom and the field and corresponds to the effect of the radiation reaction of the atom. As a result, the field operator can be written into a sum of the free part and the source part as
\beq
\mathbf{E}(x(\tau))=\mathbf{E}^f(x(\tau))+\mathbf{E}^s(x(\tau))
\eeq
with
\bea
\mathbf{E}^f(x(\tau))&=&{i\/{2\pi c^2}}\sqrt{\hbar\/{\pi\varepsilon_0}}
\int^{\infty}_0d\omega\;\omega^2\times\nn\\&&\quad\;\quad\;\quad\;\quad\;
\int d^3\textbf{r}'\sqrt{\epsilon_I(\textbf{r}',\omega)}\;
\texttt{G}(\textbf{r}(\tau),\textbf{r}',\omega)\cdot\textbf{a}(t(\tau),\textbf{r}',\omega)+\mathbf{H.C.}\;,
\label{Ef}\\
\mathbf{E}^s(x(\tau))&=&-{i\/{\hbar}}
\int^{\tau}_{\tau_0}d\tau'\;[\bm{\mu}(\tau')\cdot\mathbf{E}(x(\tau')),\textbf{E}(x(\tau))]\;,
\label{Es}
\eea
where ``H.C." denotes the Hermitian conjugate term. On the right hand side of the above two equations, we have replaced  operators $\textbf{a}^f$ and $\textbf{E}^f$ with $\textbf{a}$ and $\textbf{E}$ which is correct for the first order approximation.

Assume that
the system is composed of two half spaces, one at a  temperature $T_s$, and the other at a temperature $T_e$. Generally, $T_s$ doesn't coincide with $T_e$, and we assume that each part is in local thermal equilibrium.
For the system composed of the substrate and the environment, we denote the state of the quantum electromagnetic field with $|\beta_s,\beta_e\rangle$ in which $\beta_s={\hbar c\/{k_B T_s}}$, $\beta_e={\hbar c\/{k_B T_e}}$ and $k_B$ is the Boltzmann constant. The density operator of the state is $\rho=\rho_s\bigotimes\rho_e$ with $\rho_s=e^{-{H_F}/{k_BT_s}}$ and $\rho_e=e^{-{H_F}/{k_BT_e}}$ being the density operators of the two subsystems (the substrate and the environment) respectively.  Now with the free part and the source part given in Eqs.~(\ref{Ef}) and (\ref{Es}), we can analyze the rate of change of an arbitrary observable of the atom, $O(\tau)$, in terms of $\textbf{E}^f$ (corresponding to the effect of the thermal fluctuations) and $\textbf{E}^s$ (corresponding to the effect of radiation reaction of the atom). Following  DDC~\cite{DDC82,DDC84},  we choose the symmetric ordering between the operators of the atom and the field to identify the contributions of the thermal fluctuations and radiation reaction to the rate of change of $O(\tau)$, and we obtain
\bea
\biggl({dO(\tau)\/d\tau}\biggr)_{tf}&=&-{i\/{2\hbar}}(\textbf{E}^f(x(\tau))\cdot[\bm{\mu}(\tau),O(\tau)]
+[\bm{\mu}(\tau),O(\tau)]\cdot\textbf{E}^f(x(\tau)))\;,\\
\biggl({dO(\tau)\/d\tau}\biggr)_{rr}&=&-{i\/{2\hbar}}(\textbf{E}^s(x(\tau))\cdot[\bm{\mu}(\tau),O(\tau)]
+[\bm{\mu}(\tau),O(\tau)]\cdot\textbf{E}^s(x(\tau)))\;.
\eea
Taking the average value of the above two equations over the state of the field, $|\beta_s,\beta_e\rangle$, and proceeding in a manner similar to that in Refs.~\cite{DDC84,Audretsch95},  we can identify, in the resulting expressions, the part that acts as an effective Hamiltonian for the atomic observable, which is
\beq
\biggl\langle\beta_s,\beta_e\bigg|\biggl({dO(\tau)\/d\tau}\biggr)_{tf,rr}\bigg|\beta_s,\beta_e\biggr\rangle
=i[H^{eff}_{tf,rr},O(\tau)]+\mathrm{non-Hamiltonian\;terms}
\eeq
with
\bea
H^{eff}_{tf}(\tau)&=&-{i\/{2\hbar}}\int^{\tau}_{\tau_0}d\tau'
(C^F_{ij})_{\beta_s,\beta_e}(x(\tau),x(\tau'))[\mu_i(\tau),\mu_j(\tau')]\;,
\label{eff tf}\\
H^{eff}_{rr}(\tau)&=&-{i\/{2\hbar}}\int^{\tau}_{\tau_0}d\tau'
(\chi^F_{ij})_{\beta_s,\beta_e}(x(\tau),x(\tau'))\{\mu_i(\tau),\mu_j(\tau')\}\;,
\label{eff rr}
\eea
where $(C^F_{ij})_{\beta_s,\beta_e}(x(\tau),x(\tau'))$ and $(\chi^F_{ij})_{\beta_s,\beta_e}(x(\tau),x(\tau'))$ are respectively the symmetric correlation function and linear susceptibility function of the field defined as
\bea
(C^F_{ij})_{\beta_s,\beta_e}(x(\tau),x(\tau'))
&=&{1\/2}\langle\beta_s,\beta_e|\{\mathrm{E}_i(x(\tau)),\mathrm{E}_j(x(\tau'))\}|\beta_s,\beta_e\rangle\;,
\label{Cf}\\
(\chi^F_{ij})_{\beta_s,\beta_e}(x(\tau),x(\tau'))
&=&{1\/2}\langle\beta_s,\beta_e|[\mathrm{E}_i(x(\tau)),\mathrm{E}_j(x(\tau'))]|\beta_s,\beta_e\rangle\;.
\label{chif}
\eea
Assuming that the atom is initially in state $|a\rangle$,  and taking  the average value of Eqs.~(\ref{eff tf}) and (\ref{eff rr}) over the state, we obtain the contributions of the thermal fluctuations and radiation reaction to the energy shift of the atom respectively as
\bea
(\delta E_a)_{tf}&=&-{i\/{\hbar}}\int^{\tau}_{\tau_0}d\tau'(C^F_{ij})_{\beta_s,\beta_e}(x(\tau),x(\tau'))
(\chi^A_{ij})_a(\tau,\tau')\;,\label{tf contribution integration}\\
(\delta E_a)_{rr}&=&-{i\/{\hbar}}\int^{\tau}_{\tau_0}d\tau'(\chi^F_{ij})_{\beta_s,\beta_e}(x(\tau),x(\tau'))
(C^A_{ij})_a(\tau,\tau')\;.\label{rr contribution integration}
\eea
In the above two equations, $(\chi^A_{ij})_a(\tau,\tau')$ and $(C^A_{ij})_a(\tau,\tau')$ are two statistical functions of the atom in state $|a\rangle$ which are defined as
\bea
(\chi^A_{ij})_a(\tau,\tau')&=&{1\/2}\langle a|[\mu_i(\tau),\mu_j(\tau')]|a\rangle\;,\\
(C^A_{ij})_a(\tau,\tau')&=&{1\/2}\langle a|\{\mu_i(\tau),\mu_j(\tau')\}|a\rangle
\eea
and they can be further explicitly written as
\bea
(\chi^A_{ij})_a(\tau,\tau')&=&{1\/2}\sum_b[\langle a|\mu_i(0)|b\rangle\langle b|\mu_j(0)|a\rangle e^{i\omega_{ab}(\tau-\tau')}+\langle a|\mu_j(0)|b\rangle\langle b|\mu_i(0)|a\rangle e^{-i\omega_{ab}(\tau-\tau')}]\;,\label{atom chi}\nn\\\\
(C^A_{ij})_a(\tau,\tau')&=&{1\/2}\sum_b[\langle a|\mu_i(0)|b\rangle\langle b|\mu_j(0)|a\rangle e^{i\omega_{ab}(\tau-\tau')}-\langle a|\mu_j(0)|b\rangle\langle b|\mu_i(0)|a\rangle e^{-i\omega_{ab}(\tau-\tau')}]\label{atom c}\nn\\
\eea
where $\omega_{ab}=\omega_{a}-\omega_{b}$ and the summation extends over the complete set of the atomic states.

To evaluate the contributions of the thermal fluctuations and radiation reaction to the energy shift of the atom,  we need  the correlation functions of the field, i.e.,  Eqs.~(\ref{Cf}) and (\ref{chif}). Our next task is to find these functions. For this purpose, let us further assume that the right half-space with $z>0$ is filled with a thermal bath at a temperature $T_e$,   the left half-space is filled with a dielectric substrate at a temperature $T_s$,  each half-space is in local thermal equilibrium, and the surface of the substrate coincides with the plane $z=0$. By using the fluctuation-dissipation theorem together with the local source hypothesis~\cite{Landau63}, the two correlation functions of the field can be expressed as  (see Appendix.~\ref{app:A})
\bea
&&(C^F_{ij})^{bnd}_{\beta_s,\beta_e}(x(\tau),x(\tau'))\nn\\
&=&{\hbar\delta_{ij}\/{\pi\varepsilon_0c^2}}\int^{\infty}_0d\omega\;\omega^2
\biggl({1\/2}+{1\/{e^{\beta_e\omega/{c}}-1}}\biggr)(e^{-i\omega(t-t')}+e^{i\omega(t-t')})
\times\mathrm{Im}[\mathrm{G}_{ij}(z,\omega)]\nn\\
&+&{\hbar\delta_{ij}\/{\pi\varepsilon_0c^2}}\int^{\infty}_0d\omega\;\omega^2
\biggl({1\/{e^{\beta_s\omega/{c}}-1}}-{1\/{e^{\beta_e\omega/{c}}-1}}\biggr)(e^{i\omega(t-t')}+e^{-i\omega(t-t')})
\times g_{ij}(z,\omega)\label{concrete cfij}
\eea
where
\bea
g_{ij}(\mathbf{r},\mathbf{r}',\omega)={\omega^2\/c^2}
\int_{z_1<0} d^3\mathbf{r}_1\epsilon_I(\mathbf{r}_1,\omega)
\mathrm{G}_{ik}(\mathbf{r},\mathbf{r}_1,\omega)\mathrm{G}^{\star}_{jk}(\mathbf{r}',\mathbf{r}_1,\omega)\;,
\eea
and
\bea
(\chi^F_{ij})^{bnd}_{\beta_s,\beta_e}(x(\tau),x(\tau'))={\hbar\delta_{ij}\/{2\pi\varepsilon_0c^2}}
\int^{\infty}_0d\omega\;\omega^2(e^{-i\omega(t-t')}-e^{i\omega(t-t')})
\times\mathrm{Im}[\mathrm{G}_{ij}(z,\omega)]\;.\label{concrete chifij}
\eea

\section{Energy shift of  an atom near the surface of a general dielectric substrate}

With the field correlation functions found, now we are able to calculate the energy shift of an atom out of thermal equilibrium near the surface of a general dielectric substrate. Inserting the statistical function of the atom, Eq.~(\ref{atom chi}), and the symmetric correlation function of the field, Eq.~(\ref{concrete cfij}), into Eq.~(\ref{tf contribution integration}),  we find the contribution of the thermal fluctuations to the energy shift of the atom
\bea
(\delta E_a)^{bnd}_{tf}&=&{1\/{\pi\varepsilon_0c^2}}\sum_{b}|\langle a|\mu_i(0)|b\rangle|^2\nn\\&&\quad\times
\int^{\infty}_0d\omega\biggl({\omega^2\/{\omega+\omega_{ab}}}-{\omega^2\/{\omega-\omega_{ab}}}\biggr)
\biggl({1\/2}+{1\/{e^{\beta_e\omega/c}-1}}\biggr)\times\mathrm{Im}[\mathrm{G}_{ii}(z,\omega)]\nn\\
&+&{1\/{\pi\varepsilon_0c^2}}\sum_{b}|\langle a|\mu_i(0)|b\rangle|^2\nn\\&&\quad\times
\int^{\infty}_0d\omega\biggl({\omega^2\/{\omega+\omega_{ab}}}-{\omega^2\/{\omega-\omega_{ab}}}\biggr)
\biggl({1\/{e^{\beta_s\omega/c}-1}}-{1\/{e^{\beta_e\omega/c}-1}}\biggr)\times g_{ii}(z,\omega)\;.\nn\\
\eea
Similarly, the insertion of Eqs.~(\ref{atom c}) and (\ref{concrete chifij}) into Eq.~(\ref{rr contribution integration}) gives rise to the contribution of radiation reaction to the energy shift of the atom
\beq
(\delta E_a)^{bnd}_{rr}=-{1\/{2\pi\varepsilon_0c^2}}\sum_{b}|\langle a|\mu_i(0)|b\rangle|^2
\int^{\infty}_0d\omega\biggl({\omega^2\/{\omega+\omega_{ab}}}+{\omega^2\/{\omega-\omega_{ab}}}\biggr)
\times\mathrm{Im}[\mathrm{G}_{ii}(z,\omega)]\;.
\eeq
Adding up the above two equations, we arrive at the total energy shift of the atom in state $|a\rangle$. For simplicity, we now consider an isotropically polarizable two-level atom with its levels being $\pm{1\/2}\hbar\omega_0$, and we define  the polarizability of the atom in state $|a\rangle$  as
\beq
\alpha=\sum_i\alpha_i=\sum_{i,b}{2|\langle a|\mu_i(0)|b\rangle|^2\/{3\hbar\omega_0}}\;.
\eeq
Now we can write the total boundary-dependent energy shift into  a sum of three parts as
\beq
(\delta E_a)^{bnd}_{tot}=(\delta E_a)^{bnd}_{vac}(z)+(\delta E_a)^{bnd}_{eq}(z,\beta_e)+(\delta E_a)^{bnd}_{neq}(z,\beta_s,\beta_e)
\label{general E tot}
\eeq
with
\bea
(\delta E_a)^{bnd}_{vac}(z)&=&-{\hbar\omega_0\alpha\/{2\pi\varepsilon_0c^2}}
\int^{\infty}_0d\omega {\omega^2\/{\omega-\omega_{ab}}}\times g_1(z,\omega)\;,
\label{general E vac}\\
(\delta E_a)^{bnd}_{eq}(z,\beta_e)&=&{\hbar\omega_0\alpha\/{2\pi\varepsilon_0c^2}}
\int^{\infty}_0d\omega\biggl({\omega^2\/{\omega+\omega_{ab}}}-{\omega^2\/{\omega-\omega_{ab}}}\biggr)
{g_1(z,\omega)\/{e^{\beta_e\omega/c}-1}}\;,
\label{general E eq}\\
(\delta E_a)^{bnd}_{neq}(z,\beta_s,\beta_e)&=&{\hbar\omega_0\alpha\/{2\pi\varepsilon_0c^2}}
\int^{\infty}_0d\omega\biggl({\omega^2\/{\omega+\omega_{ab}}}-{\omega^2\/{\omega-\omega_{ab}}}\biggr)
{g_2(z,\omega)\/{e^{\beta_s\omega/c}-1}}\nn\\
&-&{\hbar\omega_0\alpha\/{2\pi\varepsilon_0c^2}}
\int^{\infty}_0d\omega\biggl({\omega^2\/{\omega+\omega_{ab}}}-{\omega^2\/{\omega-\omega_{ab}}}\biggr)
{g_2(z,\omega)\/{e^{\beta_e\omega/c}-1}}\;,
\label{general E neq}
\eea
where
\bea
g_1(z,\omega)&=&\mathrm{Im}[\mathrm{G}_{xx}(z,\omega)+\mathrm{G}_{yy}(z,\omega)+\mathrm{G}_{zz}(z,\omega)]\;,
\label{general g_1}\\
g_2(z,\omega)&=&g_{xx}(z,\omega)+g_{yy}(z,\omega)+g_{zz}(z,\omega)\;.
\label{general g_2}
\eea
Here it is obvious that the first term, $(\delta E_a)^{bnd}_{vac}(z)$, corresponds to the energy shift of the atom caused by zero-point fluctuations, the second term, $(\delta E_a)^{bnd}_{eq}(z,\beta_e)$, corresponds to the contribution of the thermal fluctuations for the system in thermal equilibrium at a temperature $T_e$, and the third term, $(\delta E_a)^{bnd}_{neq}(z,\beta_s,\beta_e)$, arises from the out of thermal equilibrium nature of the system.
When the temperature of the substrate and the environment coincides, i.e., $T_s=T_e$, the third term which reflects the revision generated by the effect out of thermal equilibrium vanishes and the result of thermal equilibrium is recovered.

Combining Eqs.~(\ref{general g_1}) and (\ref{general g_2}) with Eqs.~(\ref{green function})-(\ref{parameters}), $g_1(z,\omega)$ and $g_2(z,\omega)$ can be expressed, after lengthy simplifications, as
\beq
g_1(z,\omega)=g_{11}(z,\omega)+g_{12}(z,\omega)\label{g_1}
\eeq
with
\bea
g_{11}(z,\omega)&=&{\omega\/{4\pi c}}\int^{1}_0 dt
\biggl[{{t^2-|\epsilon-1+t^2|}\/{|t+\sqrt{\epsilon-1+t^2}|^2}}+
{{(|\epsilon|^2t^2-|\epsilon-1+t^2|)(1-2t^2)}\/{|\epsilon t+\sqrt{\epsilon-1+t^2}|^2}}\biggr]\cos(2\omega z t/c)\nn\\
&+&{\omega\/{2\sqrt{2}\pi c}}\int^{1}_0 dt\;t\sqrt{|\epsilon-1+t^2|-(\epsilon_R-1+t^2)}
\nn\\&&\quad\quad\quad\quad\times\biggl[{1\/{|t+\sqrt{\epsilon-1+t^2}|^2}}-
{{(|\epsilon-1+t^2|+t^2-1)(1-2t^2)}\/{|\epsilon t+\sqrt{\epsilon-1+t^2}|^2}}\biggr]\sin(2\omega z t/c)\;,\nn\\
\eea
\bea
g_{12}(z,\omega)&=&{\omega\/{2\sqrt{2}\pi c}}\int^{\infty}_0 dt\;t\;e^{-{2\omega z\/c}t} \sqrt{|\epsilon-1-t^2|+(\epsilon_R-1-t^2)}\;\nn\\&&\quad\;\quad\;\quad\quad\times
\biggl[{{(t^2+1+|\epsilon-1-t^2|)(2t^2+1)}\/{|i t\epsilon+\sqrt{\epsilon-1-t^2}|^2}}
+{1\/{|i t+\sqrt{\epsilon-1-t^2}|^2}}\biggr]\;,
\eea
and
\beq
g_2(z,\omega)=g_{21}(\omega)+g_{12}(z,\omega)
\eeq
with
\bea
g_{21}(\omega)&=&{\omega\/{4\sqrt{2}\pi c}}\int^{1}_0 dt \sqrt{|\epsilon-t|+(\epsilon_R-t)}
\biggl({{t+|\epsilon-t|}\/{|\epsilon\sqrt{1-t}+\sqrt{\epsilon-t}|^2}}
+{1\/{|\sqrt{1-t}+\sqrt{\epsilon-t}|^2}}\biggr)\nn\\
\label{concrete g_{21}}
\eea
It is worth noting here that the functions $g_{11}(z,\omega)$ and $g_{21}(\omega)$ give the contributions of the traveling modes of the quantum electromagnetic field and $g_{12}(z,\omega)$ describes those of the evanescent modes. Obviously, function $g_{21}(\omega)$ is independent of $z$, thus we leave it out in the following discussions as we are concerned with the boundary-dependent energy shift of the atom.

\section{Energy shift and the Casimir-Polder force of an atom near a non-dispersive dielectric substrate}

Since an analytical computation of the integrals  Eqs.~(\ref{general E tot})-(\ref{general E neq}) looks like mission impossible, we now apply the general results we derived in the preceding  section to the atom near a non-dispersive dielectric substrate with real constant relative permittivity.  Before that, we will first look at a special case, i.e, the case of a perfect conductor, which
 corresponds to an infinitely large real  relative permittivity, i.e., $\epsilon\rightarrow\infty$,  and in this case, we can deduce from Eqs.~(\ref{g_1})-(\ref{concrete g_{21}}) that
\bea
g_1(z,\omega)&=&f(z,\omega)\nn\\
            &=&-{c\/{4\pi\omega z^2}}\cos(2\omega z/c)-{1\/{4\pi z}}\sin(2\omega z/c)
               +{c^2\/{8\pi z^3\omega^2}}\sin(2\omega z/c)\;,\label{conduncting plane g1}\\
g_2(z,\omega)&=&0\;.\label{conduncting plane g2}
\eea
Combining Eq.~(\ref{conduncting plane g2}) with Eq.~(\ref{general E neq}), we find that $(\delta E_a)^{bnd}_{neq}(z,\beta_s,\beta_e)=0$ . This means that effects from being out of thermal equilibrium vanish for a perfect conductor,  and as a result the total energy shift of the atom in state $|a\rangle$ can be simplified to
\bea
(\delta E_a)^{bnd}_{tot}&=&-{\hbar\omega_0\alpha\/{2\pi\varepsilon_0c^2}}
\int^{\infty}_0d\omega {\omega^2\/{\omega-\omega_{ab}}}\times f(z,\omega)\nn\\
&&+{\hbar\omega_0\alpha\/{2\pi\varepsilon_0c^2}}
\int^{\infty}_0d\omega\biggl({\omega^2\/{\omega+\omega_{ab}}}-{\omega^2\/{\omega-\omega_{ab}}}\biggr)
{1\/{e^{\beta_e\omega/c}-1}}\times f(z,\omega)\;.
\label{E tot-conduncting plane}
\eea
This expression is in a form  different   from and a bit simpler than  that  in Ref.~\cite{Z-Y09} for an atom in a thermal bath near a conducting plane obtained using the field correlation functions found by the method of images, which involves both integration and summation over an infinite series.  We do not plan to prove mathematically that they are equivalent.  However, we will demonstrate that they do agree in the special circumstances which are examined in Ref.~\cite{Z-Y09}.
Using Eq.~(\ref{E tot-conduncting plane}), we can show that in the low temperature limit, when the wavelength of the thermal photons is much larger than the transition wavelength of the atom,
i.e., ${\beta_e\/\lambda_0}\gg1$ where $\lambda_0={c\/\omega_0}$, we have
for the ground-state atom,
\bea
(\delta E_-)^{bnd}_{tot}\approx
\left\{
    \begin{array}{ll}
     -{\hbar\/{4\pi\varepsilon_0}}[{\alpha\omega_0\/{8z^3}}+{32\pi^5\alpha cz^2\/{315\beta_e^6}}], \quad\;\; z\ll\lambda_0\ll\beta_e\;, \\
     -{\hbar\/{4\pi\varepsilon_0}}[{3\alpha c\/{8\pi z^4}}+{32\pi^5\alpha cz^2\/{315\beta_e^6}}], \;\quad\lambda_0\ll z\ll\beta_e\; ,\\
     -{\hbar\/{4\pi\varepsilon_0}}{\alpha c\/{4z^3\beta_e}}, \quad\quad\quad\quad\quad\quad\lambda_0\ll\beta_e\ll z\;,
     \end{array}
     \right.
     \label{low temp limit-conducting plane-2}
\eea
and for the excited atom,
\bea
(\delta E_+)^{bnd}_{tot}\approx
\left\{
  \begin{array}{ll}
  -{\hbar\/{4\pi\varepsilon_0}}[{\alpha\omega_0\/{8z^3}}-{32\pi^5\alpha cz^2\/{315\beta_e^6}}],\quad\quad\quad\quad
\quad\quad\quad\quad\quad\quad\quad\quad\quad\quad\quad\quad\;\; z\ll\lambda_0\ll\beta_e, \\
  {\hbar\/{4\pi\varepsilon_0}}[({\alpha\omega^3_0\/{2zc^2}}-{\alpha\omega_0\/{4z^3}})\cos({2z\omega_0\/c})
   -{\alpha\omega^2_0\/{2z^2c}}\sin({2z\omega_0\/c})+{3\alpha c\/{8\pi z^4}}+{32\pi^5\alpha cz^2\/{315\beta_e^6}}], \lambda_0\ll z\ll\beta_e ,\\
    {\hbar\/{4\pi\varepsilon_0}}[({\alpha\omega^3_0\/{2zc^2}}-{\alpha\omega_0\/{4z^3}})\cos({2z\omega_0\/c})
   -{\alpha\omega^2_0\/{2z^2c}}\sin({2z\omega_0\/c})+{\alpha c\/{4z^3\beta_e}}], \quad\quad\quad\quad \lambda_0\ll\beta_e\ll z.
  \end{array}
\right.\nn\\
\label{low temp limit-conducting plane-1}
\eea
Note that in both the short and intermediate distance regions ($z\ll\lambda_0\ll\beta_e$ and $\lambda_0\ll z\ll\beta_e$), the revision induced by thermal fluctuations to the energy shift for the atom in both the ground and excited states is proportional to $z^2T_e^6$. This seems to differ from the result in Ref.~\cite{Eberlein} in which the contribution of thermal fluctuations in the leading order is found to be proportional to $T^4$ (see Eqs.~(6.3) and (6.6) in Ref.~\cite{Eberlein}). However, these two results are actually not contradictory to each other as here we are concerned with  the distance-dependent energy shift of the atom and the $T^4$ term is distance-independent.
Similarly,
in the high temperature limit, when the wavelength of the thermal photons is much smaller than the transition wavelength of the atom, i.e., ${\beta_e\/\lambda_0}\ll1$,
we find for the ground-state atom,
\bea
(\delta E_-)^{bnd}_{tot}\approx
\left\{
  \begin{array}{ll}
  -{\hbar\/{4\pi\varepsilon_0}}[{\alpha\omega_0\/{8z^3}}-{4\pi^3\alpha\omega_0^2z^2\/{75c \beta_e^4}}],\quad\quad\quad\quad\quad\quad\quad\quad\quad\quad\quad z\ll\beta_e\ll\lambda_0, \\
  -{\hbar\/{4\pi\varepsilon_0}}[{\alpha\omega_0\/{8z^3}}-{\alpha\omega^4_0z\/{2\beta_e c^3}}], \;\quad\quad\quad\quad\quad\quad\quad\quad\quad\quad\quad\;\;\;\;\beta_e\ll z\ll\lambda_0 ,\\
   -{\hbar\/{4\pi\varepsilon_0}}[{\alpha\omega^2_0\/{2z\beta_e c}}\cos({2z\omega_0\/c})
   -{\alpha\omega_0\/{2\beta_e z^2}}\sin({2z\omega_0\/c})+{\alpha c\/{4\beta_e z^3}}], \;\;\beta_e\ll\lambda_0\ll z ,
  \end{array}
  \label{high temp limit-conducting plane-2}
\right.
\eea
and for the excited atom,
\bea
(\delta E_+)^{bnd}_{tot}\approx
\left\{
  \begin{array}{ll}
  -{\hbar\/{4\pi\varepsilon_0}}[{\alpha\omega_0\/{8z^3}}+{4\pi^3\alpha\omega_0^2z^2\/{75c \beta_e^4}}],\quad\quad\quad\quad\quad\quad\quad\quad\quad\quad\;\;z\ll\beta_e\ll\lambda_0, \\
  -{\hbar\/{4\pi\varepsilon_0}}[{\alpha\omega_0\/{8z^3}}+{\alpha\omega^4_0z\/{2\beta_e c^3}}], \quad\quad\quad\quad\quad\quad\quad\quad\quad\quad\;\;\;\;\;\;\;\beta_e\ll z\ll\lambda_0 ,\\
    {\hbar\/{4\pi\varepsilon_0}}[{\alpha\omega^2_0\/{2z\beta_e c}}\cos({2z\omega_0\/c})
   -{\alpha\omega_0\/{2\beta_e z^2}}\sin({2z\omega_0\/c})+{\alpha c\/{4\beta_e z^3}}],\;\;\;\beta_e\ll\lambda_0\ll z.
  \end{array}
  \label{high temp limit-conducting plane-1}
\right.
\eea
These results agree with those obtained in Ref.~\cite{Z-Y09} for a two-level atom near a perfect conducting plane  in interaction with quantum electromagnetic fields in a thermal bath at thermal equilibrium.


Now let us turn to the main focus of the paper, which  is the atom-wall force for a two-level atom out of thermal equilibrium  near a dielectric substrate with a real constant permittivity. In this case, the functions $g_{1}(z,\omega)$ and $g_{2}(z,\omega)$ can be simplified to
\bea
g_{1}(z,\omega)&=&{\omega\/2\pi c} \int_0^1 dt\;[2\mathrm{T}_{\parallel}(t)+\mathrm{T}_{\perp}(t)]\cos(2z\omega t/c)+g_{12}(z,\omega)\;,
\label{g_1 for real dielectric}\\
g_{2}(z,\omega)&=&g_{12}(z,\omega)
\label{g_2 for real dielectric}
\eea
where
\bea
\mathrm{A}_{\parallel}(t)&=&{1\/2}\sqrt{\epsilon-1}{{(2\epsilon+1)(\epsilon-1)t^2+1}\/{(\epsilon^2-1)t^2+1}}t\sqrt{1-t^2}\;,\label{Apara}\\
\mathrm{A}_{\perp}(t)&=&\epsilon\sqrt{\epsilon-1}{{(\epsilon-1)t^2+1}\/{(\epsilon^2-1)t^2+1}}t\sqrt{1-t^2}\;,\\
\mathrm{T}_{\parallel}(t)&=&{1\/4}\biggl({{t-\sqrt{\epsilon-1+t^2}}\/{t+\sqrt{\epsilon-1+t^2}}}-t^2{{\epsilon t-\sqrt{\epsilon-1+t^2}}\/{\epsilon t+\sqrt{\epsilon-1+t^2}}}\biggr)\;,\\
\mathrm{T}_{\perp}(t)&=&{1\/2}(1-t^2){{\epsilon t-\sqrt{\epsilon-1+t^2}}\/{\epsilon t+\sqrt{\epsilon-1+t^2}}}\;,\label{Tperp}
\eea
and
\beq
g_{12}(z,\omega)={\omega\/2\pi c}\int_{0}^{1} dt\;[2\mathrm{A}_{\parallel}(t)+\mathrm{A}_{\perp}(t)]e^{-2z\sqrt{\epsilon-1}\omega t/c}\;.
\eeq
Then by inserting Eqs.~(\ref{g_1 for real dielectric}) and (\ref{g_2 for real dielectric}) into Eqs.~(\ref{general E tot})-(\ref{general E neq}), the three parts of the energy shift of the atom in state $|a\rangle$ can now be re-expressed as
\bea
(\delta E_a)^{bnd}_{vac}(z)&=&-{\hbar\omega_0\alpha\/{4\pi^2\varepsilon_0c^3}}
\int^{\infty}_0d\omega {\omega^3\/{\omega-\omega_{ab}}}\sum_{\sigma}\mathrm{W}_\sigma f_{\sigma}(z,\omega)\;,
\label{E vac for real dielectric}\\
(\delta E_a)^{bnd}_{eq}(z,\beta_e)&=&{\hbar\omega_0\alpha\/{4\pi^2\varepsilon_0c^3}}
\int^{\infty}_0d\omega\biggl({\omega^3\/{\omega+\omega_{ab}}}-{\omega^3\/{\omega-\omega_{ab}}}\biggr){1\/{e^{\beta_e\omega/c}-1}}
\sum_{\sigma}\mathrm{W}_\sigma f_{\sigma}(z,\omega)\;,\nn\\
\label{E eq for real dielectric}\\
(\delta E_a)^{bnd}_{neq}(z,\beta_s,\beta_e)&=&{\hbar\omega_0\alpha\/{4\pi^2\varepsilon_0c^3}}
\int^{\infty}_0d\omega\biggl({\omega^3\/{\omega+\omega_{ab}}}-{\omega^3\/{\omega-\omega_{ab}}}\biggr)
\biggl({1\/{e^{\beta_s\omega/c}-1}}-{1\/{e^{\beta_e\omega/c}-1}}\biggr)\nn\\
&&\quad\quad\times\sum_{\sigma}\int^1_0dt\;\mathrm{W}_\sigma \mathrm{A}_\sigma(t)e^{-2z\sqrt{\epsilon-1}\omega t/c}\;,
\label{E neq for real dielectric}
\eea
where
\beq
f_{\sigma}(z,\omega)=\int^1_0dt\;[\mathrm{A}_\sigma(t)e^{-2z\sqrt{\epsilon-1}\omega t/c}+\mathrm{T}_\sigma(t)\cos(2z\omega t/c)]\label{function f(z,omega)}
\eeq
with $\sigma=\parallel,\perp$ and $\mathrm{W}_{\parallel}=2$, $\mathrm{W}_{\perp}=1$. The above three parts sum  to the total boundary-dependent energy shift of the atom.

Noticing the relation
\beq
{\omega^3\/{\omega-\omega_{ab}}}=\omega^2+\omega\omega_{ab}+{\omega\omega_{ab}^2\/{\omega-\omega_{ab}}}\;,
\eeq
we can divide the first part, $(\delta E_a)^{bnd}_{vac}(z)$, which corresponds to the contribution of zero-point fluctuations into a sum of three parts as
\beq
(\delta E_a)^{bnd}_{vac}(z)=(\delta E_{a})^{bnd}_{vac-1}(z)+(\delta E_{a})^{bnd}_{vac-2}(z)+(\delta E_{a})^{bnd}_{vac-3}(z)
\eeq
with
\bea
(\delta E_{a})^{bnd}_{vac-1}(z)&=&-{\hbar\omega_0\alpha\/{4\pi^2\varepsilon_0c^3}}
\int^{\infty}_0 d\omega\;\omega^2[2f_{\parallel}(z,\omega)+f_{\perp}(z,\omega)]\;,\label{Evac1}\\
(\delta E_{a})^{bnd}_{vac-2}(z)&=&-{\hbar\omega_0\alpha\omega_{ab}\/{4\pi^2\varepsilon_0c^3}}\int^{\infty}_0 d\omega\;\omega [2f_{\parallel}(z,\omega)+f_{\perp}(z,\omega)]\;,\label{Evac2}\\
(\delta E_{a})^{bnd}_{vac-3}(z)&=&-{\hbar\omega_0\alpha\omega^2_{ab}\/{4\pi^2\varepsilon_0c^3}}\int^{\infty}_0 d\omega\;{\omega\/{\omega-\omega_{ab}}} [2f_{\parallel}(z,\omega)+f_{\perp}(z,\omega)]\label{Evac3}
\eea
and then we can calculate them  one by one.
For the double-integral  in $(\delta E_{a})^{bnd}_{vac-1}(z)$,  we find, using the method proposed in Refs.~\cite{Siklos99,Eberlein99} (see  Appendix.~{\ref{app:B}}),
\bea
\mathrm{I}_{1\sigma}&=&\int^{\infty}_0 d\omega\;\omega^2 f_{\sigma}(z,\omega)\nn\\
&=&-{c^3\/8z^3}\biggl[{\pi\/2}\mathrm{T}_{\sigma}''(0)+{2\/{(\epsilon-1)^{3/2}}}
\biggl(\mathrm{A}_{\sigma}'(0)-\int^1_0dt\;{{\mathrm{A}_{\sigma}(t)-\mathrm{A}_{\sigma}'(0)t}\/t^3}\biggr)\biggr]\;.
\eea
Combining the above result with the concrete forms of $\mathrm{T}_{\sigma}(t)$ and $\mathrm{A}_{\sigma}(t)$ (see Eqs.~(\ref{Apara})-(\ref{Tperp})) yields 
\beq
(\delta E_{a})^{bnd}_{vac-1}(z)=-{{\epsilon-1}\/{\epsilon+1}}{\hbar\/{4\pi\varepsilon_0}}{\alpha\omega_0\/{8z^3}}\;.
\eeq
This term is proportional to $z^{-3}$ at an arbitrary position. Actually, it corresponds to the contribution of the electrostatic interaction in the minimal coupling scheme (see Eq.~(3.26) in Ref.~\cite{Eberlein99}).
The double-integral in Eq.~(\ref{Evac2}) has  been calculated  in Ref.~\cite{Eberlein99}, so here we just list it  without giving the details,
\bea
\mathrm{I}_{2\sigma}&=&\int^{\infty}_0 d\omega\;\omega f_{\sigma}(\omega,z)\nn\\
&=&{c^2\/4z^2}\biggl[\mathrm{T}_{\sigma}(0)-\int^1_0dt\;{{\mathrm{T}_{\sigma}(t)-\mathrm{T}_{\sigma}(0)
-{\mathrm{A}_{\sigma}(t)\/{\epsilon-1}}}\/{t^2}}+{\mathrm{A}_{\sigma}'(0)\/{\epsilon-1}}\ln\sqrt{\epsilon-1}\biggr]\;.
\eea
Putting this result into Eq.~(\ref{Evac2}), we find that $(\delta E_{a})^{bnd}_{vac-2}(z)$ is proportional to $z^{-2}$ for the atom at an arbitrary distance from the surface of the dielectric substrate. This term  corresponds to the average value of $e^2\mathbf{A}^2\/2m$ (where $\mathbf{A}$ represents the vector potential operator of the electromagnetic field) in the minimal coupling scheme, and  it is actually the self energy of an electron at a distance $z$ from the surface of the dielectric substrate.
For $(\delta E_{a})^{bnd}_{vac-3}(z)$, the double-integral  in Eq.~(\ref{Evac2}) is also discussed in Ref.~\cite{Eberlein99}. It  corresponds to the contribution of the term $-{e\/m}\mathbf{A}\cdot\mathbf{p}$ in the minimal coupling scheme, i.e., the coupling between the momentum of the electron and the vector potential of the quantum field. An exact analytical result  for an arbitrary position is however difficult to get,  but in two limiting cases, the approximate analytical results are obtainable.

In the short distance region where $\{2z,2z\sqrt{\epsilon-1}\}\ll\lambda_0$~\footnote{Hereafter, $\{a,b\}\ll c$ means $a\ll c$ and $b\ll c$. Similarly, $\{a,b\}\gg c$ means $a\gg c$ and $b\gg c$.},
the leading term of the double-integral  in $(\delta E_{a})^{bnd}_{vac-3}(z)$ is
\bea
\mathrm{I}_{3\sigma}&=&\int^{\infty}_0 d\omega\;{\omega\/{\omega-\omega_{ab}}} f_{\sigma}(\omega,z)\nn\\
&\approx&{\pi c\/4z}\mathrm{T}_{\sigma}(0)+{c\/{4z\sqrt{\epsilon-1}}}\int^1_0dt\;{{\mathrm{A}_{\sigma}(t)}\/t}\;,
\eea
yielding a  $(\delta E_{a})^{bnd}_{vac-3}(z)$  proportional to $z^{-1}$. As a result, $(\delta E_{a})^{bnd}_{vac-1}(z)$ prevails over the other terms, and we have
\beq
(\delta E_a)^{bnd}_{vac}(z)\approx(\delta E_{a})^{bnd}_{vac-1}(z)
=-{\hbar\/{4\pi\varepsilon_0}}{{\epsilon-1}\/{\epsilon+1}}{\alpha\omega_0\/{8z^3}}.
\label{contribution of vacuum short distance}
\eeq
This shows that in the short distance region, $\{2z,2z\sqrt{\epsilon-1}\}\ll\lambda_0$, no matter if the atom is in its excited state or the ground state, the boundary-dependent energy shift due to zero-point fluctuations is proportional to $z^{-3}$,  and the resulting atom-wall force obeys the van der Waals law.

In the long distance region, i.e., when  $\{2z,2z\sqrt{\epsilon-1}\}\gg\lambda_{0}$, after complicated simplifications, we find that
\bea
\mathrm{I}_{3\sigma}&\approx&
-{\mathrm{I}_{2\sigma}\/\omega_{ab}}-{\mathrm{I}_{1\sigma}\/\omega^2_{ab}}
+{\omega_{ab}c^4\/{16z^4|\omega^4_{ab}|}}g_{\sigma}(\epsilon)
+\pi\theta(\omega_{ab})\biggl[{c\mathrm{T}_{\sigma}(1)\/{2z}}\cos({2z\omega_{ab}}/c)
-{c^2\mathrm{T}'_{\sigma}(1)\/{4z^2\omega_{ab}}}\sin({2z\omega_{ab}/c})
\nn\\&&-{c^3\mathrm{T}''_{\sigma}(1)\/{8z^3\omega_{ab}^2}}\cos({2z\omega_{ab}/c})
+{c^4\mathrm{T}^{(3)}_{\sigma}(1)\/{16z^4\omega_{ab}^3}}\sin({2z\omega_{ab}/c})\biggr]\;,
\label{long distance T3rho}
\eea
where
\bea
g_{\sigma}(\epsilon)&=&2\mathrm{T}_{\sigma}(0)+3\mathrm{T}'_{\sigma}(0)+3\mathrm{T}''_{\sigma}(0)
+{{3\mathrm{A}'_{\sigma}(0)-\mathrm{A}^{(3)}_{\sigma}(0)\ln\sqrt{\epsilon-1}}\/{(\epsilon-1)^2}}\nn\\&&
-6\int^1_0dt\;{{\mathrm{T}_{\sigma}(t)-\mathrm{T}_{\sigma}(0)-\mathrm{T}'_{\sigma}(0)t-{\mathrm{T}''_{\sigma}(0)\/2}t^2
+{{\mathrm{A}_{\sigma}(t)-\mathrm{A}'_{\sigma}(0)t}\/{(\epsilon-1)^2}}}\/{t^4}}\label{g-epsilon}
\eea
and $\theta(\omega_{ab})$ is the step-function defined as
\beq
\theta(\omega_{ab})=\left\{
                      \begin{array}{ll}
                        1, \quad\;\omega_{ab}>0\;, \\
                        0, \quad\;\omega_{ab}<0\;.
                      \end{array}
                    \right.
\eeq
For the details on how to get Eq.~(\ref{long distance T3rho}), see Ref.~\cite{Eberlein99}. Here we point out that in the expression of Eq.~(B31) in Ref.~\cite{Eberlein99}, there is a typo for the sign of the fourth term in the coefficient of the term $\varsigma^{-4}$ (concerning the expression of $g(\epsilon)$ here) and we have corrected it. A substitution of Eq.~(\ref{long distance T3rho})  into Eq.~(\ref{Evac3}) reveals that for the ground-state atom ($\omega_{ab}<0$)$, (\delta E_{a})^{bnd}_{vac-3}(z)$ is proportional to $z^{-2}$ as the leading term $\mathrm{I}_{2\sigma}\propto z^{-2}$, while  for the excited atom ($\omega_{ab}>0$), $(\delta E_{a})^{bnd}_{vac-3}(z)$ oscillates with the distance between the atom and the surface of the substrate. Adding up the three parts, we find that $(\delta E_{a})^{bnd}_{vac-1}(z)$ and $(\delta E_{a})^{bnd}_{vac-2}(z)$ are completely canceled by parts of $(\delta E_{a})^{bnd}_{vac-3}(z)$, and as a result, the boundary-dependent energy shift due to zero-point fluctuations in the long distance region becomes
\bea
(\delta E_-)^{bnd}_{vac}(z)&\approx&{\hbar\/{4\pi\varepsilon_0}}{\alpha c\/{16\pi z^4}}g(\epsilon)\;,
\label{vac-long distance E-}\\
(\delta E_+)^{bnd}_{vac}(z)&\approx&-{\hbar\/{4\pi\varepsilon_0}}\biggl[{{1-\sqrt{\epsilon}}\/{1+\sqrt{\epsilon}}}
\biggl({\alpha\omega_0^3\/2zc^2}\cos(2z\omega_0/c)-{\alpha\omega_0^2\/2z^2c}\sin(2z\omega_0/c)\biggr)+{\alpha c\/{16\pi z^4}}g(\epsilon)\biggr] \label{vac-long distance E+}\nn\\
\eea
with
\bea
g(\epsilon)&=&2g_{\parallel}(\epsilon)+g_{\perp}(\epsilon)\nn\\
&=&{{-6\epsilon^2+3\epsilon^{3/2}+4\epsilon+3\sqrt{\epsilon}-10}\/{\epsilon-1}}
+{{3(2\epsilon^3-4\epsilon^2+3\epsilon+1)}\/{(\epsilon-1)^{3/2}}}\ln[\sqrt{\epsilon}+\sqrt{\epsilon-1}]\nn\\&&
+{{6\epsilon^2}\/{\sqrt{\epsilon+1}}}\ln\biggl[{{1+\sqrt{\epsilon+1}}\/{\epsilon+\sqrt{\epsilon(\epsilon+1)}}}\biggr]\;.
\eea
As is shown in the following figure, for $\epsilon>1$, $g(\epsilon)$ is always negative.
\begin{figure}[!htb]
\centering
\label{file}
\includegraphics[scale=0.65]{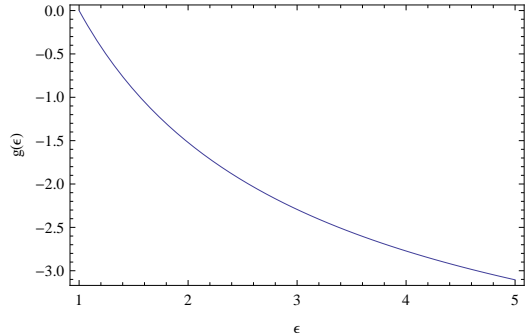}
\caption{$g(\epsilon)$ for $\epsilon\geq 1$.}
\end{figure}
Thus for the ground-state atom, $(\delta E_-)^{bnd}_{vac}(z)$ is proportional to $z^{-4}$ and is always negative, leading to an attractive  Casimir-Polder force proportional to $z^{-5}$. For the excited atom, $(\delta E_+)^{bnd}_{vac}(z)$ usually oscillates with the distance between the atom and the surface of the dielectric substrate, and the amplitude of  oscillation is much larger than that of the constant term  proportional to $z^{-4}$, thus $(\delta E_+)^{bnd}_{vac}(z)$ can be either positive or negative or can even  be zero.  Accordingly, the Casimir-Polder force due to the zero-point fluctuations can be either repulsive or attractive or can even  be zero. Let us note that  the above result is not valid for the case of a perfect conducting plane in which $\epsilon\rightarrow\infty$. In this case, we should be careful in taking the limit  of $\epsilon\rightarrow\infty$.  In fact,  we should take the limit $\epsilon\rightarrow\infty$ in $\mathrm{T}_{\sigma}(t)$ and $\mathrm{A}_{\sigma}(t)$ before performing differentiation  on them when simplifying Eq.~(\ref{g-epsilon}).  Then by so doing, we find that $g_{\parallel}(\epsilon)=g_{\perp}(\epsilon)=-2$, and
\bea
(\delta E_-)^{bnd}_{tot}(z)&\approx&-{\hbar\/4\pi\varepsilon_0}{3\alpha\/8\pi z^4}\;,\label{E_-vac-long-perfect conducting plane}\\
(\delta E_+)^{bnd}_{tot}(z)&\approx&{\hbar\/4\pi\varepsilon_0}\biggl[{\alpha\omega_0^3\/2zc^2}\cos(2z\omega_0/c)
-{\alpha\omega_0^2\/2z^2c}\sin(2z\omega_0/c)+{3\alpha c\/8\pi z^4}\biggr]\;.
\eea
Thereby, the energy shift of an isotropically polarizable two-level atom  far from the surface of a perfect conducting plane is recovered.

Until now we have only discussed the atomic energy shift and the Casimir-Polder force due to zero-point fluctuations. Next,  we will
turn our attention to the contributions of the thermal fluctuations.  It is difficult to get  analytical results for a general case. However,  fortunately,
we are able to find asymptotic behaviors  in the low and high temperature limits.
In the following discussion, we assume that the difference between the temperature of the substrate, $T_s$, and that of the environment, $T_e$, is neither extremely large nor extremely small.

\subsection{Low temperature limit }

We first deal with the low temperature limit,  i.e., ${\beta_s\/\lambda_0}\gg1$ and ${\beta_e\/\lambda_0}\gg1$. For simplicity, we abbreviate these two conditions by $\{\beta_s,\beta_e\}\gg\lambda_0$. Here, we will analyze  how the energy shift and the atom-wall force behave as the distance varies.  Since now we have two length scales, i.e., the transition wavelength of the atom $\lambda_0$ and  the wavelength of thermal photons $\beta_s$ or $\beta_e$, we can define a short distance region  where  $\{2z,2z\sqrt{\epsilon-1}\}\ll\lambda_0\ll\{\beta_s,\beta_e\}$. By doing the $\omega$-integration before the $t$-integration, $(\delta E_{\pm})^{bnd}_{eq}(z,\beta_e)$ can be simplified to
\bea
(\delta E_-)^{bnd}_{eq}(z,\beta_e)&=&-(\delta E_+)^{bnd}_{eq}(z,\beta_e)\nn\\
&\approx&{\hbar\/{4\pi\varepsilon_0}}\biggl[{96\zeta[5]\/\pi}{c\alpha z\/\beta_e^5}f_1(\epsilon)+{16\pi^5\/63}{c\alpha z^2\/\beta_e^6}f_2(\epsilon)\biggr]\;,
\label{eq contribution-real permittivity}
\eea
with
\bea
f_1(\epsilon)&=&{{\pi(\epsilon-1)(3\epsilon^3+11\epsilon^2+\epsilon+1)}\/{16(\epsilon+1)^3}}\;,\\
f_2(\epsilon)&=&\int^{1}_0 dt\;t^2
\biggl[{{1-\epsilon}\/{(t+\sqrt{\epsilon-1+t^2})^2}}+
{{(1-2t^2)((\epsilon^2-1)t^2-(\epsilon-1))}\/{(\epsilon t+\sqrt{\epsilon-1+t^2})^2}}\biggr]\;.
\eea
Here it is easy to see that the term proportional to $z\beta_e^{-5}$ in Eq.~(\ref{eq contribution-real permittivity}) which is absent in the case of a perfect conducting plane (see Eqs.~(\ref{low temp limit-conducting plane-2}) and (\ref{low temp limit-conducting plane-1})) is much larger than that proportional to $z^2\beta_e^{-6}$.

For the effect out of thermal equilibrium (Eq.~(\ref{E neq for real dielectric})), similarly, we find, when $2z\sqrt{\epsilon-1}\ll\{\beta_s,\beta_e\}$,
that
\bea
(\delta E_-)^{bnd}_{neq}(z,\beta_s,\beta_e)&=&-(\delta E_+)^{bnd}_{neq}(z,\beta_s,\beta_e)\nn\\
&\approx&{\hbar\/{4\pi\varepsilon_0}}\biggl[{96\zeta[5]\/\pi}{c\alpha z\/\beta_s^5}f_1(\epsilon)-{96\zeta[5]\/\pi}{c\alpha z\/\beta_e^5}f_1(\epsilon)\biggr]\;.
\label{neq contribution-real permittivity}
\eea
Here it is worth pointing out that Eqs.~(\ref{eq contribution-real permittivity}) and (\ref{neq contribution-real permittivity}) are not valid for the case of a perfect conducting plane as they are obtained under the conditions, $\{2z,2z\sqrt{\epsilon-1}\}\ll\beta_e$ and $2z\sqrt{\epsilon-1}\ll\{\beta_s,\beta_e\}$ respectively, which means that the parameter $\epsilon$ can not be infinitely large.
Adding up Eqs.~(\ref{eq contribution-real permittivity}) and (\ref{neq contribution-real permittivity}) gives rise to the total boundary-dependent energy shift of the excited and ground-state atoms due to the thermal fluctuations
\bea
(\delta E_-)^{bnd}_{ther}(z,\beta_s,\beta_e)&=&-(\delta E_+)^{bnd}_{ther}(z,\beta_s,\beta_e)\nn\\
&\approx&{\hbar\/{4\pi\varepsilon_0}}\biggl[{96\zeta[5]\/\pi}{c\alpha z\/\beta_s^5}f_1(\epsilon)+{16\pi^5\/63}{c\alpha z^2\/\beta_e^6}f_2(\epsilon)\biggr]\;.\label{low-temp-short-intermediate-thermal}
\eea
Notice that this result is valid in the region $\{2z,2z\sqrt{\epsilon-1}\}\ll\{\beta_s,\beta_e\}$.
One can see that although both the thermal fluctuations associated with the substrate and the environment contribute to the atomic energy shift  in this region,  the contribution of the former dominates over the latter.

Combining the above result for the contribution of the thermal fluctuations with  the contribution of zero-point fluctuations, Eq.~(\ref{contribution of vacuum short distance}), we find that in the short distance region, $\{2z,2z\sqrt{\epsilon-1}\}\ll\lambda_0\ll\{\beta_s,\beta_e\}$,  the total boundary-dependent energy shift for an isotropically polarizable  two-level atom in the stationary regime out of thermal equilibrium is
\bea
(\delta E_-)^{bnd}_{tot}(z)&\approx&-{\hbar\/{4\pi\varepsilon_0}}\biggl[{{\epsilon-1}\/{\epsilon+1}}{\alpha\omega_0\/{8z^3}}
-{96\zeta[5]\/\pi}{c\alpha z\/\beta_s^5}f_1(\epsilon)-{16\pi^5\/63}{c\alpha z^2\/\beta_e^6}f_2(\epsilon)\biggr]\;,\\
(\delta E_+)^{bnd}_{tot}(z)&\approx&-{\hbar\/{4\pi\varepsilon_0}}\biggl[{{\epsilon-1}\/{\epsilon+1}}{\alpha\omega_0\/{8z^3}}
+{96\zeta[5]\/\pi}{c\alpha z\/\beta_s^5}f_1(\epsilon)+{16\pi^5\/63}{c\alpha z^2\/\beta_e^6}f_2(\epsilon)\biggr]\;.
\eea
Obviously,  the thermal fluctuations associated with  both the substrate and the environment contribute to the atomic energy shift. Comparing the contribution due to the thermal fluctuations with that due to zero-point fluctuations  characterized by the term proportional to $z^{-3}$, we find that the revision caused by the thermal fluctuations is negligible. Thus the Casimir-Polder force the atoms in both the ground state and the excited state feel is attractive and proportional to $z^{-4}$ which is the van der Waals law.

We can also introduce an intermediate distance region where $\lambda_0\ll\{2z,2z\sqrt{\epsilon-1}\}\ll\{\beta_s,\beta_e\}$, then we find, by
combining Eq.~(\ref{low-temp-short-intermediate-thermal}) with the contributions of zero-point fluctuations, Eqs.~(\ref{vac-long distance E-}) and (\ref{vac-long distance E+})), that in this region
\bea
(\delta E_-)^{bnd}_{tot}(z)&\approx&{\hbar\/{4\pi\varepsilon_0}}\biggl[{\alpha c\/{16\pi z^4}}g(\epsilon)
+{96\zeta[5]\/\pi}{c\alpha z\/\beta_s^5}f_1(\epsilon)+{16\pi^5\/63}{c\alpha z^2\/\beta_e^6}f_2(\epsilon)\biggr]\;,
\label{realdie-lowtemp-interdis-E-}\\
(\delta E_+)^{bnd}_{tot}(z)&\approx&-{\hbar\/{4\pi\varepsilon_0}}
\biggl[{{1-\sqrt{\epsilon}}\/{1+\sqrt{\epsilon}}}\biggl({\alpha\omega_0^3\/2zc^2}\cos(2z\omega_0/c)
-{\alpha\omega_0^2\/2z^2c}\sin(2z\omega_0/c)\biggr)+{\alpha c\/{16\pi z^4}}g(\epsilon)
\nn\\&&\quad\;\quad\;\quad+{96\zeta[5]\/\pi}{c\alpha z\/\beta_s^5}f_1(\epsilon)
+{16\pi^5\/63}{c\alpha z^2\/\beta_e^6}f_2(\epsilon)\biggr]\;.
\eea
Similarly, as in the short distance region,  the thermal fluctuations associated with both the substrate and the environment contribute to the atomic energy shift. For the ground-state atom, the contribution due to the thermal fluctuations is much smaller than that due to zero-point fluctuations  characterized by the term proportional to $z^{-4}$, and so the Casimir-Polder force the atom feels is attractive (as $g(\epsilon)<0$) and proportional to $z^{-5}$ . For the excited atom, oscillatory  terms caused by zero-point fluctuations appear and the amplitude of oscillation is much larger than the terms due to the thermal fluctuations and the term proportional to $z^{-4}$.  As a result,  the atomic energy shift can be either negative or positive or can even  be zero, yielding an atom-wall force that can be either  attractive or repulsive or can even be zero.

Finally let us turn to the long distance region where $\lambda_0\ll\{\beta_s,\beta_e\}\ll\{2z,2z\sqrt{\epsilon-1}\}$.
When $\{2z,2z\sqrt{\epsilon-1}\}\gg\beta_e$,  $(\delta E_{\pm})^{bnd}_{eq}(z,\beta_e)$ can be calculated by performing the integrations in Eq.~(\ref{E eq for real dielectric}) (see Appendix.~{\ref{app:C}) to get
\bea
(\delta E_-)^{bnd}_{eq}(z,\beta_e)&=&-(\delta E_+)^{bnd}_{eq}(z,\beta_e)\nn\\
&\approx&-{\hbar\/{4\pi\varepsilon_0}}{\alpha c\/{4\beta_e z^3}}{{\epsilon-1}\/{\epsilon+1}}\;,
\label{eq for real dielectric-long distance}
\eea
and when $2z\sqrt{\epsilon-1}\gg\{\beta_s,\beta_e\}$,  treating Eq.~(\ref{E neq for real dielectric}) in a similar way (see Appendix.~{\ref{app:C}) leads to
\bea
(\delta E_-)^{bnd}_{neq}(z,\beta_s,\beta_e)&=&-(\delta E_+)^{bnd}_{neq}(z,\beta_s,\beta_e)\nn\\
&\approx&-{\hbar\/{4\pi\varepsilon_0}}{\pi\alpha c\/{12z^2}}{{\epsilon+1}\/{\sqrt{\epsilon-1}}}\biggl({1\/\beta_s^2}-{1\/\beta_e^2}\biggr)\;.
\label{neq for real dielectric-long distance}
\eea
Notice that in Eq.~(\ref{eq for real dielectric-long distance}), if we take the limit $\epsilon\rightarrow\infty$, we recover the contribution of the thermal fluctuations to the energy shift of an atom at a distance $z$ from a perfect conducting plane in a thermal bath at a  temperature $T_e$ (see the third line in both Eqs.~(\ref{low temp limit-conducting plane-2}) and Eq.~(\ref{low temp limit-conducting plane-1})). But trouble appears if we take the $\epsilon\rightarrow\infty$ limit in Eq.~(\ref{neq for real dielectric-long distance}) as the result would be divergent. However, as pointed out in the paragraph above Eq.~(\ref{E_-vac-long-perfect conducting plane}), we should take the limit $\epsilon\rightarrow\infty$ in $\mathrm{T}_{\sigma}(t)$ and $\mathrm{A}_{\sigma}(t)$ before taking their derivatives. Then following steps as those taken in Appendix.~{\ref{app:C}, we get
\bea
(\delta E_-)^{bnd}_{neq}(z,\beta_s,\beta_e)&=&-(\delta E_+)^{bnd}_{neq}(z,\beta_s,\beta_e)\nn\\
&\approx&-{\hbar\/{4\pi\varepsilon_0}}{\pi\alpha c\/{4z^2\sqrt{\epsilon-1}}}\biggl({1\/\beta_s^2}-{1\/\beta_e^2}\biggr)\nn\\
&\approx&0
\eea
which means that for the perfect conducting plane, the effect of non-thermal equilibrium vanishes due to the infinite $\epsilon$.

For a general real dielectric substrate, adding Eqs.~(\ref{eq for real dielectric-long distance}) and (\ref{neq for real dielectric-long distance}), we obtain the total contribution of the thermal fluctuations to the atomic energy shift. Under the assumption that the temperature of the substrate, $T_s$, and that of the environment, $T_e$, are not extremely close, the result can be approximated by
\bea
(\delta E_-)^{bnd}_{ther}(z,\beta_s,\beta_e)&=&-(\delta E_+)^{bnd}_{ther}(z,\beta_s,\beta_e)\nn\\
&\approx&(\delta E_-)^{bnd}_{neq}(z,\beta_s,\beta_e)\nn\\
&\approx&-{\hbar\/{4\pi\varepsilon_0}}{\pi\alpha c\/{12z^2}}{{\epsilon+1}\/{\sqrt{\epsilon-1}}}\biggl({1\/\beta_s^2}-{1\/\beta_e^2}\biggr)\;,
\label{low-temp-thermal long distance}
\eea
since $(\delta E_\pm)^{bnd}_{eq}(z,\beta_e)$ is negligible as compared to $(\delta E_\pm)^{bnd}_{neq}(z,\beta_s,\beta_e)$.  This result is valid in the region where $\{2z,2z\sqrt{\epsilon-1}\}\gg\{\beta_s,\beta_e\}$. So, in this region, the contribution of the effect of non-thermal equilibrium to the atomic energy shift prevails over the effect of thermal equilibrium. Noteworthily,
here both contributions of the thermal fluctuations of the substrate and that of the environment are of the same order and are all proportional to $z^{-2}$ but with opposite signs.
It is then a matter of an easy differentiation exercise to  get the Casimir-Polder force due to the thermal fluctuations
\bea
(F_-)^{bnd}_{ther}(z,\beta_s,\beta_e)&=&-(F_+)^{bnd}_{ther}(z,\beta_s,\beta_e)\nn\\
&\approx&-{\hbar\/{4\pi\varepsilon_0}}{\pi\alpha c\/{6z^3}}{{\epsilon+1}\/{\sqrt{\epsilon-1}}}\biggl({1\/\beta_s^2}-{1\/\beta_e^2}\biggr)\;.
\eea
Thus, for an atom in its ground (excited) state, the Casimir-Polder force is attractive (repulsive) if the temperature of the substrate, $T_s$, is higher than that of the environment, $T_e$, and repulsive (attractive) if otherwise. Here it is worth pointing out that our result for the ground-state atom is consistent with that obtained by M. Antezza, et al. in Refs. \cite{Antezza05,Antezza06} (see Eq.~(12) in Ref. \cite{Antezza05}) although the issue is dealt with from a different perspective in the present paper. Moreover, in Refs.~\cite{Antezza05,Antezza06}, the result is obtained by mathematically assuming $z\rightarrow\infty$, and thus the physical region where this result is valid is not clearly given. In contrast, here we find out the concrete region.
Notice that we use  SI units  while the Gauss units are adopted in Refs. \cite{Antezza05,Antezza06}, so a discrepancy of a factor $({4\pi\varepsilon_0})^{-1}$ appears between our results and theirs.

Combining  Eq.~(\ref{low-temp-thermal long distance}), with the contributions of zero-point fluctuations, Eqs.~(\ref{vac-long distance E-}) and (\ref{vac-long distance E+}), we find that in the long distance region, $\lambda_0\ll\{\beta_s,\beta_e\}\ll\{2z,2z\sqrt{\epsilon-1}\}$, the total boundary-dependent energy shift of the  atom is
\bea
(\delta E_-)^{bnd}_{tot}(z)&\approx&-{\hbar\/{4\pi\varepsilon_0}}\biggl[{{\epsilon+1}\/{\sqrt{\epsilon-1}}}{\pi\alpha c\/{12z^2}}\biggl({1\/\beta_s^2}-{1\/\beta_e^2}\biggr)-{\alpha c\/{16\pi z^4}}g(\epsilon)\biggr]\;,
\label{realdie-lowtemp-longdis-E-}\\
(\delta E_+)^{bnd}_{tot}(z)&\approx&-{\hbar\/{4\pi\varepsilon_0}}
\biggl[{{1-\sqrt{\epsilon}}\/{1+\sqrt{\epsilon}}}\biggl({\alpha\omega_0^3\/2zc^2}\cos(2z\omega_0/c)
-{\alpha\omega_0^2\/2z^2c}\sin(2z\omega_0/c)\biggr)\nn\\&&\quad\;\quad\;\quad
-{{\epsilon+1}\/{\sqrt{\epsilon-1}}}{\pi\alpha c\/{12z^2}}\biggl({1\/\beta_s^2}-{1\/\beta_e^2}\biggr)+{\alpha c\/{16\pi z^4}}g(\epsilon)\biggr]\;.
\eea
Notice that in this region, if $T_s$ and $T_e$ are not extremely close,  the contribution of the effect of non-thermal equilibrium for the energy shift of the ground-state atom dominates over the contribution of zero-point fluctuations which is proportional to $z^{-4}$, thus the Casimir-Polder force it feels behaves like $(\mathrm{T}_s^2-\mathrm{T}_e^2)/z^3$. If $T_s>T_e$, the force is attractive and it is repulsive  otherwise. For the excited atom, as the amplitude of the oscillatory terms  is always larger than the contribution of the effect of non-thermal equilibrium and the term proportional to $z^{-4}$, the energy shift of the atom can be either  negative or positive and can even be zero. As a result, the Casimir-Polder force for the excited atom  can be either  attractive or repulsive or can even be zero.

\subsection{High temperature limit}

We now analyze the behavior of the atom-wall force out of thermal equilibrium in the high temperature limit, i.e., when${\beta_s\/\lambda_0}\ll1$ and ${\beta_e\/\lambda_0}\ll1$, which is not considered in Ref.~\cite{Antezza05,Antezza06}. We can combine these conditions into $\{\beta_s,\beta_e\}\ll\lambda_0$.  We then find in the short-distance region where $\{2z,2z\sqrt{\epsilon-1}\}\ll\beta_e\ll\lambda_0$,
\bea
(\delta E_-)^{bnd}_{eq}(z,\beta_e)&=&-(\delta E_+)^{bnd}_{eq}(z,\beta_e)\nn\\
&\approx&-{\hbar\/{4\pi\varepsilon_0}}\biggl[8\zeta[3]\cdot{\alpha\omega_0^2 z\/\pi c\beta_e^3}f_1(\epsilon)+{2\pi^3\alpha\omega_0^2z^2\/15 c\beta_e^4}\cdot(f_2(\epsilon)-f_3(\epsilon))\biggr]
\label{eq contribution-real permittivity-high temp-short}
\eea
with
\beq
f_3(\epsilon)=2(\epsilon-1)^{3/2}\int^1_0dt\;t^3\sqrt{1-t^2}{{(3\epsilon^2-2\epsilon-1)t^2+(\epsilon+1)}\/{(\epsilon^2-1)t^2+1}}\;.
\eeq
Here the term proportional to $z\beta_e^{-3}$ which is absent  in the case of a conducting plane (see Eqs.~(\ref{high temp limit-conducting plane-2}) and (\ref{high temp limit-conducting plane-1}))  dominates over the term proportional to $z^2\beta_e^{-4}$.
When $2z\sqrt{\epsilon-1}\ll\{\beta_s,\beta_e\}\ll\lambda_0$, we can show that
\bea
(\delta E_-)^{bnd}_{neq}(z,\beta_s,\beta_e)&=&-(\delta E_+)^{bnd}_{neq}(z,\beta_s,\beta_e)\nn\\
&\approx&-{\hbar\/{4\pi\varepsilon_0}}\biggl[8\zeta[3]\cdot{\alpha\omega_0^2 z\/\pi c}\biggl({{1}\/{\beta_s^3}}-{{1}\/{\beta_e^3}}\biggr)f_1(\epsilon)\nn\\&&\quad\quad\quad
-{2\pi^3\alpha\omega_0^2z^2\/15 c}\biggl({1\/{\beta_s^4}}-{1\/{\beta_e^4}}\biggr)f_3(\epsilon)\biggr]\;.
\label{neq contribution-real permittivity-high temp-short}
\eea
For details on how to get the above analytical result, see Appendix.~{\ref{app:D}.
By adding  Eqs.~(\ref{eq contribution-real permittivity-high temp-short}) and (\ref{neq contribution-real permittivity-high temp-short}), the total contributions of the thermal fluctuations to the boundary-dependent energy shift of the ground-state and excited atoms out of thermal equilibrium are found to be
\bea
(\delta E_-)^{bnd}_{ther}(z,\beta_s,\beta_e)&=&-(\delta E_+)^{bnd}_{ther}(z,\beta_s,\beta_e)\nn\\
&\approx&-{\hbar\/{4\pi\varepsilon_0}}\biggl[8\zeta[3]\cdot{\alpha\omega_0^2z\/\pi c\beta_s^3}f_1(\epsilon)+{2\pi^3\alpha\omega_0^2z^2\/15 c\beta_e^4}f_2(\epsilon)
-{2\pi^3\alpha\omega_0^2z^2\/15 c\beta_s^4}f_3(\epsilon)\biggr]\nn\\
&\approx&-{\hbar\/{4\pi\varepsilon_0}}\biggl[8\zeta[3]\cdot{\alpha\omega_0^2z\/\pi c\beta_s^3}f_1(\epsilon)+{2\pi^3\alpha\omega_0^2z^2\/15 c\beta_e^4}f_2(\epsilon)\biggr]\;.\nn\\
\eea
Notice that this result is valid in the short-distance region where $\{2z,2z\sqrt{\epsilon-1}\}\ll\{\beta_s,\beta_e\}\ll\lambda_0$.
Just as in the case of the low temperature limit,  the thermal fluctuations that originate from both the substrate and the environment contribute to the atomic energy shift and the former (characterized by $z\beta_s^{-3}$) is much larger than the latter (characterized by $z^2\beta_e^{-4}$).

Combining the above result with the contributions of zero-point fluctuations, Eq.~(\ref{contribution of vacuum short distance}), gives rises to the total boundary-dependent energy shift of the atom in the short distance region, $\{2z,2z\sqrt{\epsilon-1}\}\ll\{\beta_s,\beta_e\}\ll\lambda_0$,
\bea
(\delta E_-)^{bnd}_{tot}(z)&\approx&-{\hbar\/{4\pi\varepsilon_0}}\biggl[{{\epsilon-1}\/{\epsilon+1}}{\alpha\omega_0\/{8z^3}}
+8\zeta[3]\cdot{\alpha\omega_0^2z\/\pi c\beta_s^3}f_1(\epsilon)+{2\pi^3\alpha\omega_0^2z^2\/15 c\beta_e^4}f_2(\epsilon)\biggr]\;,\\
(\delta E_+)^{bnd}_{tot}(z)&\approx&-{\hbar\/{4\pi\varepsilon_0}}\biggl[{{\epsilon-1}\/{\epsilon+1}}{\alpha\omega_0\/{8z^3}}
-8\zeta[3]\cdot{\alpha\omega_0^2z\/\pi c\beta_s^3}f_1(\epsilon)-{2\pi^3\alpha\omega_0^2z^2\/15 c\beta_e^4}f_2(\epsilon)\biggr]\;.
\eea
Obviously, in this region, the contribution of zero-point fluctuations  characterized by the term proportional to $z^{-3}$ prevails over the contribution of the thermal fluctuations, thus the Casimir-Polder force  is attractive and proportional to $z^{-4}$ no matter if the atom is in its ground-state or the excited state.

Now let us look at the intermediate distance region where $\{\beta_s,\beta_e\}\ll\{2z,2z\sqrt{\epsilon-1}\}\ll\lambda_0$. In this region, we have
\bea
(\delta E_-)^{bnd}_{eq}(z,\beta_e)&=&-(\delta E_+)^{bnd}_{eq}(z,\beta_e)\nn\\
&\approx&{\hbar\/{4\pi\varepsilon_0}}{{\alpha\omega_0^2}\/{4c\beta_e z}}f_4(\epsilon)
\label{eq contribution-real permittivity-high temp-intermediate}
\eea
with
\beq
f_4(\epsilon)={{(3\epsilon+1)(\epsilon-1)}\/(\epsilon+1)^2}\;,
\eeq
and
\bea
(\delta E_-)^{bnd}_{neq}(z,\beta_s,\beta_e)&=&-(\delta E_+)^{bnd}_{neq}(z,\beta_s,\beta_e)\nn\\
&\approx&{\hbar\/{4\pi\varepsilon_0}}{{\alpha\omega_0^2}\/{4cz}}\biggl({1\/\beta_s}-{1\/\beta_e}\biggr)f_5(\epsilon)
\label{neq contribution-real permittivity-high temp-intermediate}
\eea
with
\beq
f_5(\epsilon)={{(5\epsilon+2)\epsilon+1}\/(\epsilon+1)^2}\;.
\eeq
This  shows that for the ground-state atom, the force is repulsive (attractive) if $T_s>T_e$ ($T_s<T_e$), and it is the other way around for the excited atom.
Adding Eq.~(\ref{eq contribution-real permittivity-high temp-intermediate}) and Eq.~(\ref{neq contribution-real permittivity-high temp-intermediate}), we get the total contribution of the thermal fluctuations to the boundary-dependent energy shift of the atom
\bea
(\delta E_-)^{bnd}_{ther}(z,\beta_s,\beta_e)&=&-(\delta E_+)^{bnd}_{ther}(z,\beta_s,\beta_e)\nn\\
&\approx&{\hbar\/{4\pi\varepsilon_0}}{{\alpha\omega_0^2}\/{4cz}}\biggl({f_5(\epsilon)\/\beta_s}-{2\/\beta_e}\biggr)\;.
\label{high-temp intermediate distance thermal}
\eea
Again,   the thermal fluctuations of  both the substrate and that of the environment contribute to the boundary-dependent energy shift of the atom out of thermal equilibrium, but now their contributions are of the same order and are all proportional to $z^{-1}$.
Combining Eq.~(\ref{high-temp intermediate distance thermal}) with the contributions of zero-point fluctuations, Eq.~(\ref{contribution of vacuum short distance}), yields the total boundary-dependent energy shift of the atom
\bea
(\delta E_-)^{bnd}_{tot}(z)&\approx&-{\hbar\/{4\pi\varepsilon_0}}\biggl[{{\epsilon-1}\/{\epsilon+1}}{\alpha\omega_0\/{8z^3}}
-{{\alpha\omega_0^2}\/{4cz}}\biggl({f_5(\epsilon)\/\beta_s}-{2\/\beta_e}\biggr)\biggr]\;,\\
(\delta E_+)^{bnd}_{tot}(z)&\approx&-{\hbar\/{4\pi\varepsilon_0}}\biggl[{{\epsilon-1}\/{\epsilon+1}}{\alpha\omega_0\/{8z^3}}
+{{\alpha\omega_0^2}\/{4cz}}\biggl({f_5(\epsilon)\/\beta_s}-{2\/\beta_e}\biggr)\biggr]\;.
\eea
Thus, for the ground-state (excited) atom, if ${T_s\/T_e}<{2\/f_5(\epsilon)}$ (${T_s\/T_e}>{2\/f_5(\epsilon)}$), the boundary-dependent energy shift is negative and the Casimir-Polder force  on the atom is attractive, and if $T_s f_5(\epsilon)-2T_e<{{\epsilon-1}\/{\epsilon+1}}{c\/2z^2\omega_0}$ ($T_s f_5(\epsilon)-2T_e>{{\epsilon-1}\/{\epsilon+1}}{c\/2z^2\omega_0}$), the boundary-dependent energy shift is negative (positive), and thus the Casimir-Polder force is attractive (repulsive).

Finally, let us turn our attention to the long distance region where $\{2z,2z\sqrt{\epsilon-1}\}\gg\beta_e\gg\lambda_0$. For a finite $\epsilon$, we find
\bea
(\delta E_-)^{bnd}_{eq}(z,\beta_e)&=&-(\delta E_+)^{bnd}_{eq}(z,\beta_e)\nn\\
&\approx&-{\hbar\/{4\pi\varepsilon_0}}{{\alpha\omega_0^2}\/{2\beta_e cz}}f_6(\epsilon)\cos(2z\omega_0/c)
\label{eq-real dielectric-high temp-long distance}
\eea
with
\beq
f_6(\epsilon)={{\sqrt{\epsilon}-1}\/{\sqrt{\epsilon}+1}}\;,
\eeq
and for an infinite $\epsilon$ which corresponds to the case of a perfect conducting plane,  we find by following the same procedure as that  in the case of  the low temperature limit,
\bea
(\delta E_-)^{bnd}_{eq}(z,\beta_e)&=&-(\delta E_+)^{bnd}_{eq}(z,\beta_e)\nn\\
&\approx&-{\hbar\/{4\pi\varepsilon_0}}\biggl[{\alpha\omega^2_0\/{2\beta_e z c}}\cos({2z\omega_0/c})
   -{\alpha\omega_0\/{2\beta_e z^2}}\sin({2z\omega_0/c})+{\alpha c\/{4\beta_e z^3}}\biggr]\;.
\label{perfect conductor-high temp-long distance}
\eea
which is exactly the same  as the result in Eq.~(\ref{high temp limit-conducting plane-1}).
Similarly, we find,  in the region $2z\sqrt{\epsilon-1}\gg\{\beta_s,\beta_e\}\gg\lambda_0$, that for a finite $\epsilon$,
\bea
(\delta E_-)^{bnd}_{neq}(z,\beta_s,\beta_e)&=&-(\delta E_+)^{bnd}_{neq}(z,\beta_s,\beta_e)\nn\\
&\approx&{\hbar\/{4\pi\varepsilon_0}}{\alpha c\/{4z^3}}\biggl({1\/\beta_s}-{1\/\beta_e}\biggr)f_7(\epsilon)
\label{neq-real dielectric-high temp-long distance}
\eea
with
\beq
f_7(\epsilon)={{\epsilon^3-\epsilon^2+3\epsilon+1}\/{\epsilon^2-1}}\;,
\eeq
and for $\epsilon\rightarrow\infty$,
\bea
(\delta E_-)^{bnd}_{neq}(z,\beta_s,\beta_e)&=&-(\delta E_+)^{bnd}_{neq}(z,\beta_s,\beta_e)\nn\\
&\approx&{\hbar\/{4\pi\varepsilon_0}}{3\alpha c\/{4z^3(\epsilon-1)}}\biggl({1\/\beta_s}-{1\/\beta_e}\biggr)\nn\\
&\approx&0\;,
\eea
which shows that the contribution of the effect of non-thermal equilibrium vanishes for a perfect conducting plane as expected.
Adding Eq.~(\ref{eq-real dielectric-high temp-long distance}) and  Eq.~(\ref{neq-real dielectric-high temp-long distance}), we get the contributions of the thermal fluctuations to the boundary-dependent energy shift of the atom (for finite $\epsilon$),
\bea
(\delta E_-)^{bnd}_{ther}(z,\beta_s,\beta_e)&=&-(\delta E_+)^{bnd}_{ther}(z,\beta_s,\beta_e)\nn\\
&\approx&-{\hbar\/{4\pi\varepsilon_0}}\biggl[{\alpha \omega_0^2\/{2\beta_e c z}}f_6(\epsilon)\cos(2z\omega_0/c)
-{\alpha c\/{4z^3}}\biggl({1\/\beta_s}-{1\/\beta_e}\biggr)f_7(\epsilon)\biggr]\;.
\label{high-temp-long-distance-tot-thermal}
\eea
So, in this region, the contribution of the thermal fluctuations to the atomic boundary-dependent energy shift oscillates with the distance between the atom and the surface of the substrate, and the amplitude of oscillation is always much larger than the term proportional to $z^{-3}$ if the temperature of the substrate, $T_s$, is not much higher than that of the environment, $T_e$.

For the case of a perfect conducting plane, the effect of non-thermal equilibrium vanishes, so the total contribution of the thermal fluctuations to the energy shift of the atom is actually described by Eq.~(\ref{perfect conductor-high temp-long distance}).

Combining Eq.~(\ref{high-temp-long-distance-tot-thermal}), with the contributions of zero-point fluctuations, Eqs.~(\ref{vac-long distance E-}) and (\ref{vac-long distance E+}), we obtain the total boundary-dependent energy shift of the atom  in the long distance region and in the high temperature limit
\bea
(\delta E_-)^{bnd}_{tot}(z)&\approx&-{\hbar\/{4\pi\omega_0}}\biggl[{\alpha \omega_0^2\/{2\beta_e c z}}f_6(\epsilon)\cos(2z\omega_0/c)
-{\alpha c\/{4z^3}}\biggl({1\/\beta_s}-{1\/\beta_e}\biggr)f_7(\epsilon)-{\alpha c\/{16\pi z^4}}g(\epsilon)\biggr]\;,\nn\\\\
(\delta E_+)^{bnd}_{tot}(z)&\approx&-{\hbar\/{4\pi\omega_0}}\biggl[
{{1-\sqrt{\epsilon}}\/{1+\sqrt{\epsilon}}}\biggl({\alpha\omega_0^3\/2zc^2}\cos(2z\omega_0/c)
-{\alpha\omega_0^2\/2z^2c}\sin(2z\omega_0/c)\biggr)\nn\\&&\quad\;\quad\;\quad-{\alpha \omega_0^2\/{2\beta_e c z}}f_6(\epsilon)\cos(2z\omega_0/c)+{\alpha c\/{4z^3}}\biggl
({1\/\beta_s}-{1\/\beta_e}\biggr)f_7(\epsilon)+{\alpha c\/{16\pi z^4}}g(\epsilon)\biggr]\;.\nn\\
\eea
In this region, as $T_s$ and $T_e$ are not extremely close, the term proportional to ${T_s-T_e}\/z^3$ which exists when thermal equilibrium is not reached is always much larger than the term proportional to $z^{-4}$ due to zero-point fluctuations. For the ground-state atom, the amplitude of the oscillation term due to the thermal fluctuations at equilibrium is always much larger than the second term which arises because of non-thermal equilibrium, and as a result, the boundary-dependent energy shift of the atom can be either negative or positive or  can even be zero, thus  resulting in a  Casimir-Polder force that can be either attractive or repulsive or can even be zero. For the excited atom, the energy shift and Casimir-Polder force also exhibits similar behaviors.

Let us now comment on the contributions of  the evanescent  modes from the substrate and traveling modes from the environment  to the Casimir-Polder force.  By adding  Eqs.~(\ref{E eq for real dielectric}) and (\ref{E neq for real dielectric}), it is easy for us to see that both the evanescent modes from the substrate and the traveling modes from the environment generally contribute to the atomic energy shift. In the short distance region in both the low- and high- temperature limits, the contribution of the evanescent modes from the substrate dominates over that  of the traveling modes from the environment. This conclusion also holds for an atom in the intermediate distance region and in the low temperature limit. However, for an atom  in the intermediate distance region and in the high temperature limit, the contributions of the evanescent modes from the substrate and the traveling modes from the environment are always of the same order, and the same is true  for an atom  in the long distance region in both the low- and high- temperature limits.

The above discussions are about the energy shift and Casimir-Polder force of an atom out of thermal equilibrium near the surface of a real dielectric substrate. Extending the present discussion to a general dispersive dielectric substrate for which the dielectric constant depends on the frequency, i.e., $\epsilon=\epsilon(\omega)$, the Drude model for a metal for example, is an interesting topic for future research.

\section{summary}

We have generalized the DDC formalism originally established  for  thermal equilibrium to the case  out of thermal equilibrium but in a stationary state by adopting the local source hypothesis and then we applied it to the calculation of the energy shift and the Casimir-Polder force of an atom out of thermal equilibrium near a dielectric substrate. In particular, we have calculated  the energy shift and the Casimir-Polder force of an isotropically polarizable two-level atom near  a real dielectric half-space substrate and analyzed in detail their behaviors
in three different distance regions in both the low-temperature limit and the high-temperature limit for both the ground-state and excited-state atoms.

In the low-temperature limit where the wavelength of thermal photons is assumed to be much larger than the transition wavelength of the atom, we find that in all distance regions, i.e.,  the short, intermediate and long distance regions, the thermal fluctuations that originate from both the substrate and  from the environment  contribute to the atomic energy shift and the Casimir-Polder force. In the short and intermediate distance regions, the contribution of the former is much larger than the contribution of the latter, whereas in the long distance region, the contributions of both thermal fluctuations are of the same order but with opposite signs. More importantly,   the  out of thermal equilibrium fluctuations give rise to an atom-wall force in the long distance region with a slower dependence on the distance and strong dependence on the temperature as opposed to the Lifshitz law at thermal equilibrium. In particular,  for the ground state atom, the force behaves like $(T_s^{2}-T_e^{2})/z^{3}$.  Our result in the long distance region at low temperature not only confirms that by Antezza etal obtained in a different context~\cite{Antezza05,Antezza06}, but also gives a  concrete region not clearly quantified in Refs.~\cite{Antezza05,Antezza06} where the new asymptotic behavior is valid. In the low temperature limit, the effects from being out of thermal equilibrium only become appreciable in the long distance region, while they are negligible in the short and intermediate distance regions, leading to an atom-wall force which respectively obeys the van de Waals law and the Casimir-Polder law for the ground state atoms.

In the high-temperature limit where the wavelength of thermal photons is assumed to be much smaller than the transition wavelength of the atom,  the contribution of zero-point fluctuations characterized by the term proportional to $z^{-3}$ prevails over the contribution of the thermal fluctuations in the short distance region, thus the Casimir-Polder force is attractive and proportional to $z^{-4}$ no matter if the atom is in its ground-state or the excited state.  In the intermediate distance region, the contribution of the thermal fluctuations may become comparable
to that of the zero-point fluctuations and the Casimir-Polder force may be attractive or repulsive  depending on several factors including whether the atom is the ground or excited states and the relative temperature between the substrate and the environment. Only in the long distance region do the effects of the thermal fluctuations both at  and out of thermal equilibrium  dominate over that of the zero-point fluctuations, and in this region, even the atom-wall force on the
ground state atoms becomes oscillatory around zero, meaning that the force can either be attractive or repulsive.

\begin{acknowledgments}

This work was supported in part by the NSFC under Grants No. 11075083, No. 11375092 and No. 11435006,  the SRFDP under Grant No. 20124306110001, the Zhejiang Provincial Natural Science Foundation of China under Grant No. LQ14A050001, the Research Program of Ningbo University under No. E00829134702, No. xkzwl10 and No. XYL14029, and K.C. Wong Magna Fund in Ningbo University.

\end{acknowledgments}

\appendix

\section{ Correlation functions of the field out of thermal equilibrium}
\label{app:A}
In order to find the two correlation functions of the field out of thermal equilibrium  defined in Eqs.~(\ref{Cf}) and (\ref{chif}), $(C^F_{ij})_{\beta_s,\beta_e}(x(\tau),x(\tau'))$ and $(\chi^F_{ij})_{\beta_s,\beta_e}(x(\tau),x(\tau'))$, we firstly consider the quantity
\beq
\langle \mathrm{E}_i(x(\tau)),\mathrm{E}_j(x(\tau'))\rangle_{\beta_s,\beta_e}=
\langle\beta_s,\beta_e|\mathrm{E}_i(x(\tau)),\mathrm{E}_j(x(\tau'))|\beta_s,\beta_e\rangle\;.
\eeq
Taking the Fourier transformation (see Eq.~(\ref{fourier transformation})) for the electromagnetic field operator, we can expand the above quantity  into a sum of four parts as
\bea
\langle \mathrm{E}_i(t,\mathbf{r}),\mathrm{E}_j(t',\mathbf{r}'))\rangle_{\beta_s,\beta_e}
&=&\int^{\infty}_0d\omega\int^{\infty}_0d\omega'e^{-i(\omega t-\omega't')}\langle \mathrm{E}_i(\mathbf{r},\omega)\mathrm{E}^{\dag}_j(\mathbf{r}',\omega')\rangle_{\beta_s,\beta_e}\nn\\
&+&\int^{\infty}_0d\omega\int^{\infty}_0d\omega'e^{-i(\omega t+\omega't')}\langle \mathrm{E}_i(\mathbf{r},\omega)\mathrm{E}_j(\mathbf{r}',\omega')\rangle_{\beta_s,\beta_e}\nn\\
&+&\int^{\infty}_0d\omega\int^{\infty}_0d\omega'e^{i(\omega t-\omega't')}\langle \mathrm{E}^{\dag}_i(\mathbf{r},\omega)\mathrm{E}_j(\mathbf{r}',\omega')\rangle_{\beta_s,\beta_e}\nn\\
&+&\int^{\infty}_0d\omega\int^{\infty}_0d\omega'e^{i(\omega t+\omega't')}\langle \mathrm{E}^{\dag}_i(\mathbf{r},\omega)\mathrm{E}^{\dag}_j(\mathbf{r}',\omega')\rangle_{\beta_s,\beta_e}
\label{field half corre}
\eea
where we have denoted $x(\tau)$ with $x(\tau)=(t(\tau),\mathbf{r}(\tau))$. To obtain the above equation, we have used the relation $\mathrm{E}_i(\mathbf{r},-\omega)=\mathrm{E}^{\dag}_i(\mathbf{r},\omega)$. By resorting to Eq.~(\ref{E-G-f}), we obtain
\bea
\langle \mathrm{E}_i(\mathbf{r},\omega)\mathrm{E}^{\dag}_j(\mathbf{r}',\omega')\rangle_{\beta_s,\beta_e}
&=&{\hbar\/{\pi\varepsilon_0}}{\omega^2\omega'^2\/c^4}\int d^3\mathbf{r}_1\int d^3\mathbf{r}_2
\sqrt{\epsilon_I(\mathbf{r}_1,\omega)\epsilon_I(\mathbf{r}_2,\omega')}\nn\\
&&\quad\quad \times \mathrm{G}_{ik}(\mathbf{r},\mathbf{r}_1,\omega)\mathrm{G}^{\star}_{jl}(\mathbf{r}',\mathbf{r}_2,\omega)
\langle a_k(\mathbf{r}_1,\omega)a^{\dag}_l(\mathbf{r}_2,\omega')\rangle_{\beta_s,\beta_e}
\eea
where the symbol ``$\star$'' denotes the complex conjugate. Noticing that the density operator of the thermal baths with temperatures $T_s$ and $T_e$ are separately $\rho_s=e^{-{H_F}/{k_BT_s}}$ and $\rho_e=e^{-{H_F}/{k_BT_e}}$, we find
\beq
\langle a_k(\mathbf{r}_1,\omega)a^{\dag}_l(\mathbf{r}_2,\omega')\rangle_{\beta_s,\beta_e}
=\delta_{kl}\delta(\mathbf{r}_1-\mathbf{r}_2)\delta(\omega-\omega')[1+N(\omega,T(\mathbf{r}_1))]
\eeq
with
\beq
N(\omega,T(\mathbf{r}_1))={1\/{e^{\hbar\omega/{k_BT(\mathbf{r}_1)}}}-1}
=\left\{
     \begin{array}{ll}
       {1\/{e^{\hbar\omega/{k_BT_e}}}-1}, z_1>0\;, \\
       {1\/{e^{\hbar\omega/{k_BT_s}}}-1}, z_1<0\;.
     \end{array}
   \right.
\eeq
Thus,
\bea
\langle \mathrm{E}_i(\mathbf{r},\omega)\mathrm{E}^{\dag}_j(\mathbf{r}',\omega')\rangle_{\beta_s,\beta_e}
&=&{\hbar\/{\pi\varepsilon_0}}{\omega^2\omega'^2\/c^4}\delta(\omega-\omega')\nn\\&\times&\biggl[\int_{z_1<0} d^3\mathbf{r}_1\epsilon_I(\mathbf{r}_1,\omega)
 \mathrm{G}_{ik}(\mathbf{r},\mathbf{r}_1,\omega)\mathrm{G}^{\star}_{jk}(\mathbf{r}',\mathbf{r}_1,\omega)
\biggl(1+{1\/{e^{\beta_s\omega/{c}}}-1}\biggr)\nn\\
&+&\;\int_{z_1>0} d^3\mathbf{r}_1\epsilon_I(\mathbf{r}_1,\omega)
 \mathrm{G}_{ik}(\mathbf{r},\mathbf{r}_1,\omega)\mathrm{G}^{\star}_{jk}(\mathbf{r}',\mathbf{r}_1,\omega)
\biggl(1+{1\/{e^{\beta_e\omega/{c}}}-1}\biggr)\biggr]\;.\nn\\
\eea

Similarly, we can find the average values in the other three terms in Eq.~(\ref{field half corre}), and then we have
\bea
&&\langle \mathrm{E}_i(r,\mathbf{r}),\mathrm{E}_j(t',\mathbf{r}'))\rangle_{\beta_s,\beta_e}\nn\\
&=&{\hbar\/{\pi\varepsilon_0c^4}}\int^{\infty}_0d\omega\;\omega^4 e^{-i\omega(t-t')}
\biggl(1+{1\/{e^{\beta_s\omega/c}}-1}\biggr)
\int_{z_1<0}d^3\mathbf{r}_1\epsilon_I(\mathbf{r}_1,\omega)
\mathrm{G}_{ik}(\mathbf{r},\mathbf{r}_1,\omega)\mathrm{G}^{\star}_{jk}(\mathbf{r}',\mathbf{r}_1,\omega)\nn\\
&+&{\hbar\/{\pi\varepsilon_0c^4}}\int^{\infty}_0d\omega\;\omega^4e^{-i\omega(t-t')}
\biggl(1+{1\/{e^{\beta_e\omega/c}}-1}\biggr)
\int_{z_1>0}d^3\mathbf{r}_1\epsilon_I(\mathbf{r}_1,\omega)
\mathrm{G}_{ik}(\mathbf{r},\mathbf{r}_1,\omega)\mathrm{G}^{\star}_{jk}(\mathbf{r}',\mathbf{r}_1,\omega)\nn\\
&+&{\hbar\/{\pi\varepsilon_0c^4}}\int^{\infty}_0d\omega\;\omega^4e^{i\omega(t-t')}
{1\/{e^{\beta_s\omega/c}}-1}
\int_{z_1<0}d^3\mathbf{r}_1\epsilon_I(\mathbf{r}_1,\omega)
\mathrm{G}^{\star}_{ik}(\mathbf{r},\mathbf{r}_1,\omega)\mathrm{G}_{jk}(\mathbf{r}',\mathbf{r}_1,\omega)\nn\\
&+&{\hbar\/{\pi\varepsilon_0c^4}}\int^{\infty}_0d\omega\;\omega^4e^{i\omega(t-t')}
{1\/{e^{\beta_e\omega/c}}-1}
\int_{z_1>0}d^3\mathbf{r}_1\epsilon_I(\mathbf{r}_1,\omega)
\mathrm{G}^{\star}_{ik}(\mathbf{r},\mathbf{r}_1,\omega)\mathrm{G}_{jk}(\mathbf{r}',\mathbf{r}_1,\omega)\;.
\eea
Using the relation~\cite{Dung} (see Eq.~(27))
\bea
{\omega^2\/c^2}
\int d^3\mathbf{r}_1\epsilon_I(\mathbf{r}_1,\omega)
\mathrm{G}_{ik}(\mathbf{r},\mathbf{r}_1,\omega)\mathrm{G}^{\star}_{jk}(\mathbf{r}',\mathbf{r}_1,\omega)
=\mathrm{Im}[\mathrm{G}_{ij}(\mathbf{r},\mathbf{r}',\omega)]\;,
\eea
where $\mathrm{Im}[\mathrm{G}_{ij}(\mathbf{r},\mathbf{r}',\omega)]$ represents the imaginary part of $\mathrm{G}_{ij}(\mathbf{r},\mathbf{r}',\omega)$, we deduce that
\bea
&&{\omega^2\/c^2}
\int_{z_1>0} d^3\mathbf{r}_1\epsilon_I(\mathbf{r}_1,\omega)
\mathrm{G}_{ik}(\mathbf{r},\mathbf{r}_1,\omega)\mathrm{G}^{\star}_{jk}(\mathbf{r}',\mathbf{r}_1,\omega)\nn\\
&=&\mathrm{Im}[\mathrm{G}_{ij}(\mathbf{r},\mathbf{r}',\omega)]-{\omega^2\/c^2}
\int_{z_1<0} d^3\mathbf{r}_1\epsilon_I(\mathbf{r}_1,\omega)
\mathrm{G}_{ik}(\mathbf{r},\mathbf{r}_1,\omega)\mathrm{G}^{\star}_{jk}(\mathbf{r}',\mathbf{r}_1,\omega)\;.
\eea
So $\langle \mathrm{E}_i(t,\mathbf{r}),\mathrm{E}_j(t',\mathbf{r}))\rangle_{\beta_s,\beta_e}$ can be simplified to be
\bea
&&\langle \mathrm{E}_i(t,\mathbf{r}),\mathrm{E}_j(t',\mathbf{r}'))\rangle_{\beta_s,\beta_e}\nn\\
&=&{\hbar\/{\pi\varepsilon_0c^2}}\int^{\infty}_0d\omega\;\omega^2
\biggl[\biggl(1+{1\/{e^{\beta_e\omega/{c}}-1}}\biggr)e^{-i\omega(t-t')}
+{1\/{e^{\beta_e\omega/{c}}-1}}e^{i\omega(t-t')}\biggr]\times\mathrm{Im}[\mathrm{G}_{ij}(\mathbf{r},\mathbf{r}',\omega)]\nn\\
&+&{\hbar\/{\pi\varepsilon_0c^2}}\int^{\infty}_0d\omega\;\omega^2
\biggl({1\/{e^{\beta_s\omega/{c}}-1}}-{1\/{e^{\beta_e\omega/{c}}-1}}\biggr)(e^{i\omega(t-t')}+e^{-i\omega(t-t')})
\times g_{ij}(\mathbf{r},\mathbf{r}',\omega)\nn\\
\eea
where
\bea
g_{ij}(\mathbf{r},\mathbf{r}',\omega)={\omega^2\/c^2}
\int_{z_1<0} d^3\mathbf{r}_1\epsilon_I(\mathbf{r}_1,\omega)
\mathrm{G}_{ik}(\mathbf{r},\mathbf{r}_1,\omega)\mathrm{G}^{\star}_{jk}(\mathbf{r}',\mathbf{r}_1,\omega)\;.
\eea
For an atom at $\mathbf{r}=\mathbf{r}'=(0,0,z)$, combining Eqs.~(\ref{green function})-(\ref{parameters}),
we deduce that $\mathrm{Im}[\mathrm{G}_{ij}(\mathbf{r},\mathbf{r}',\omega)]=\mathrm{Im}[\mathrm{G}_{ij}(z,\omega)]$ and $g_{ij}(\mathbf{r},\mathbf{r}',\omega)=g_{ij}(z,\omega)$  are nonzero  only when $i\neq j$ .

Using  procedures  similar to those above,  we can get $\langle \mathrm{E}_j(t',\mathbf{r}),\mathrm{E}_i(t,\mathbf{r}))\rangle_{\beta_s,\beta_e}$. So, the two correlation functions of the field can be simplified as
\bea
&&(C^F_{ij})_{\beta_s,\beta_e}(x(\tau),x(\tau'))\nn\\
&=&{\hbar\delta_{ij}\/{\pi\varepsilon_0c^2}}\int^{\infty}_0d\omega\;\omega^2
\biggl({1\/2}+{1\/{e^{\beta_e\omega/{c}}-1}}\biggr)(e^{-i\omega(t-t')}+e^{i\omega(t-t')})
\times\mathrm{Im}[\mathrm{G}_{ij}(z,\omega)]\nn\\
&+&{\hbar\delta_{ij}\/{\pi\varepsilon_0c^2}}\int^{\infty}_0d\omega\;\omega^2
\biggl({1\/{e^{\beta_s\omega/{c}}-1}}-{1\/{e^{\beta_e\omega/{c}}-1}}\biggr)(e^{i\omega(t-t')}+e^{-i\omega(t-t')})
\times g_{ij}(z,\omega)\nn\\
\eea
and
\bea
(\chi^F_{ij})_{\beta_s,\beta_e}(x(\tau),x(\tau'))={\hbar\delta_{ij}\/{2\pi\varepsilon_0c^2}}
\int^{\infty}_0d\omega\;\omega^2(e^{-i\omega(t-t')}-e^{i\omega(t-t')})
\times\mathrm{Im}[\mathrm{G}_{ij}(z,\omega)]\;.
\eea
Here we point out that in the above two correlation functions we have renormalized the term, $\mathrm{Im}[\mathrm{G}^0_{ij}(\mathbf{r},\mathbf{r}',\omega)]$, which corresponds to the fluctuations of a vacuum and is infinitely large for $\mathbf{r}=\mathbf{r}'$,  by simply subtracting it out.

\section{The double-integral in Eq.~(\ref{Evac1})}
\label{app:B}
We use here the method proposed by C. Eberlein, et. al to calculate the double-integration in Eq.~(\ref{Evac1}).
The double-integral $\mathrm{I}_{1\sigma}$ is the sum of the following two integrals,
\bea
\mathrm{I}_{1\sigma}^\mathrm{T}&=&\int^{\infty}_0d\omega\int^1_0dt\;\omega^2\mathrm{T}_\sigma(t)\cos(\eta\omega t)\;,\\
\mathrm{I}_{1\sigma}^\mathrm{A}&=&\int^{\infty}_0d\omega\int^1_0dt\;\omega^2\mathrm{A}_\sigma(t)\;e^{-\bar{\eta}\omega t}
\eea
where $\eta=2z/c$, $\bar{\eta}=\eta\sqrt{\epsilon-1}$. As the two integrals  are not separately convergent, we replace the upper-limit of the $\omega$-integral in each by a positive $\Omega$ and take it to be infinity in the end.

For $\mathrm{I}_{1\sigma}^\mathrm{T}$, if we do  the $t$-integration by parts, we get
\beq
\mathrm{I}_{1\sigma}^\mathrm{T}=-{\mathrm{T}_\sigma(1)\/\eta^2}\Omega\cos(\eta\Omega)+{\mathrm{T}_\sigma(1)\/\eta^3}\sin(\eta\Omega)
-{1\/\eta}\int^{\Omega}_0d\omega\int^1_0dt\;\omega\mathrm{T}_\sigma'(t)\sin(\eta\omega t)\;.\label{I1T-1}
\eeq
For the last term in the above equation, we can subtract the term $\mathrm{T}_\sigma'(0)$ from the $t$-integration and then add it later, i.e.,
\bea
&&{1\/\eta}\int^{\infty}_0d\omega\int^1_0dt\;\omega\mathrm{T}'_\sigma(t)\sin(\eta\omega t)\nn\\
&=&{1\/\eta}\int^{\Omega}_0d\omega\int^1_0dt\;\omega[\mathrm{T}'_\sigma(t)-\mathrm{T}'_\sigma(0)]\sin(\eta\omega t)+{\mathrm{T}'_\sigma(0)\/\eta}\int^{\Omega}_0d\omega\int^1_0dt\;\omega\sin(\eta\omega t)\;.
\eea
For the first term on the right hand side  of the above equation, we do the $t$-integration by parts, and for the second term,   wedo the double-integration directly, then we get
\bea
&&{1\/\eta}\int^{\infty}_0d\omega\int^1_0dt\;\omega\mathrm{T}'_\sigma(t)\sin(\eta\omega t)\nn\\
&=&-{\mathrm{T}'_\sigma(1)\/\eta^3}\sin(\eta\Omega)+{\mathrm{T}'_\sigma(0)\/\eta^2}\Omega
+{1\/\eta^2}\int^{\infty}_0d\omega\int^1_0dt\mathrm{T}''_\sigma(t)\cos(\eta\omega t)\;.\label{I1T-2}
\eea
Similarly, for the last term on the right  hand side of the above equation, we repeat the above steps and we get
\bea
{1\/\eta^2}\int^{\infty}_0d\omega\int^1_0dt\;\mathrm{T}''_\sigma(t)\cos(\eta\omega t)\approx{\pi\/2}{\mathrm{T''_\sigma(0)\/\eta^3}}\;.\label{I1T-3}
\eea
To obtain the above result, we have discarded the terms proportional to or of order higher than $\Omega^{-1}$. Thus
\beq
\mathrm{I}_{1\sigma}^\mathrm{T}=-{\mathrm{T}_\sigma(1)\/\eta^2}\Omega\cos(\eta\Omega)+{\mathrm{T}'_\sigma(1)\/\eta^3}\sin(\eta\Omega)
+{\mathrm{T}_\sigma(1)\/\eta^3}\sin(\eta\Omega)-{\mathrm{T}'_\sigma(0)\/\eta^2}\Omega-{\pi\/2}{\mathrm{T}''_\sigma(0)\/\eta^3}\;.
\label{re-I1T}
\eeq
Take similar steps on $\mathrm{I}_{1\sigma}^{\mathrm{A}}$  and we find
\beq
\mathrm{I}_{1\sigma}^\mathrm{A}={\mathrm{A}'_\sigma(0)\/{\bar{\eta}^2}}\Omega-{2\/{\bar{\eta}^3}}\biggl(\mathrm{A}'_\sigma(0)
-\int^1_0dt\;{{\mathrm{A}_\sigma(t)-\mathrm{A}'_\sigma(0)t}\/{t^3}}\biggr)\;.
\label{re-I1A}
\eeq
Now adding Eq.~(\ref{re-I1T}) to  Eq.~(\ref{re-I1A}) and discarding the infinite oscillating terms, we arrive at
\beq
\mathrm{I}_{1_\sigma}=-{c^3\/8z^3}\biggl[{\pi\/2}\mathrm{T}''_\sigma(0)+{2\/{(\epsilon-1)^{3/2}}}\biggl(\mathrm{A}'_\sigma(0)
-\int^1_0dt{{\mathrm{A}_\sigma(t)-\mathrm{A}'_\sigma(0)t}\/t^3}\biggr)\biggr]\;.
\eeq
Notice that to obtain the above result, we have used the relation \cite{Eberlein}
\beq
\mathrm{T}'_\sigma(0)={\mathrm{A}'_\sigma(0)\/{\epsilon-1}}\;.\label{T'-A' relation}
\eeq

\section{ Integrals in Equations.~(\ref{E eq for real dielectric}) and (\ref{E neq for real dielectric}) in the long distance region and in the low temperature limit }
\label{app:C}
The integrals in Eqs.~(\ref{E eq for real dielectric}) and (\ref{E neq for real dielectric}) are of the following forms
\bea
\mathrm{\tilde{I}}_{1\sigma}&=&\int^{\infty}_0d\omega\int^1_0dt\;\mathrm{A}_\sigma(t)
\biggl({\omega^3\/{\omega+\omega_{0}}}-{\omega^3\/{\omega-\omega_{0}}}\biggr){e^{-\bar{\eta}\omega t}\/{e^{\beta\omega/c}-1}}\;,
\label{real dielectric I_1}\\
\mathrm{\tilde{I}}_{2\sigma}&=&\int^{\infty}_0d\omega\int^1_0dt\;\mathrm{T}_\sigma(t)
\biggl({\omega^3\/{\omega+\omega_{0}}}-{\omega^3\/{\omega-\omega_{0}}}\biggr){\cos(\eta\omega t)\/{e^{\beta\omega/c}-1}}\;.
\label{real dielectric I_2}
\eea
In the low temperature limit, ${\beta\/\lambda_0}\gg1$ where $\lambda_0={c\/\omega_0}$, the above integrals can be approximated as
\bea
\mathrm{\tilde{I}}_{1\sigma}&\approx&{2c^4\/{\beta^4\omega_0}}\int^{\infty}_0dy\int^1_0dt\;\mathrm{A}_\sigma(t){y^3e^{-ayt}\/{e^y-1}}\;,
\label{I_1-1}\\
\mathrm{\tilde{I}}_{2\sigma}&\approx&{2c^4\/{\beta^4\omega_0}}\int^{\infty}_0dy\int^1_0dt\;\mathrm{T}_\sigma(t){{y^3\cos(byt)}\/{e^y-1}}
\label{I_2-1}
\eea
with $a={2z\sqrt{\epsilon-1}\/\beta}$ and $b={2z\/\beta}$.
For $\mathrm{\tilde{I}}_{1\sigma}$, we perform the $y$-integration by parts and we obtain
\beq
\mathrm{\tilde{I}}_{1\sigma}={2c^4\/{\beta^4\omega_0a}}\int^1_0dt\;{\mathrm{A}_\sigma(t)\/t}\int^{\infty}_0dy{{3y^2(e^y-1)-y^3e^y}\/{(e^y-1)^2}}e^{-ayt}\;.
\eeq
This integral can be done by subtracting $\mathrm{A}'_{\sigma}(0)$ from the $t$-integration and adding it later. Then taking the limit $a\gg1$, we get
\beq
\mathrm{\tilde{I}}_{1\sigma}\approx{2c^4\/{\beta^4\omega_0}}\biggl[{\pi^2\/6}{\mathrm{A}'_\sigma(0)\/a^2}
+{2\/a^3}\biggl(\int^1_0dt{{\mathrm{A}_\sigma(t)-\mathrm{A}'_\sigma(0)t}\/{t^3}}-\mathrm{A}'_\sigma(0)\biggr)\biggr]\;.
\label{final I_1 in low temp and long distance}
\eeq

Similarly, for $\mathrm{\tilde{I}}_{2\sigma}$, do the $t$-integration by parts and we obtain
\beq
\mathrm{\tilde{I}}_{2\sigma}\approx{2c^4\/{\beta^4\omega_0}}\biggl[{\mathrm{T}_{\sigma}(1)\/b}\int^{\infty}_0dy\;{y^2\sin(by)\/{e^y-1}}
-{1\/b}\int^{\infty}_0dy\;\int^1_0dt\;{y^2\sin(b y t)\/{e^y-1}}\mathrm{T}'_\sigma(t)\biggr]\;.
\label{I_2-2}
\eeq
We do the $t$-integration in  the second integral on the right hand side  of the above equation by parts and we obtain
\bea
\mathrm{\tilde{I}}_{2\sigma}&\approx&{2c^4\/{\beta^4\omega_0}}\biggl[{\mathrm{T}_{\sigma}(1)\/b}\int^{\infty}_0dy\;{y^2\sin(by)\/{e^y-1}}+
{\mathrm{T}'_{\sigma}(1)\/b^2}\int^{\infty}_0dy\;{y\cos(by)\/{e^y-1}}\nn\\&&\quad\;\quad\;-{\pi^2\/6}{\mathrm{T}'_{\sigma}(0)\/b^2}
-{1\/b^2}\int^{\infty}_0dy\int^1_0dt\;{y\cos(byt)\/{e^y-1}}\mathrm{T}''_{\sigma}(t)\biggr]\;.
\label{I_2-3}
\eea
For the last integral in the above square bracket, we can first subtract $\mathrm{T}''_{\sigma}(0)$ from the t-integration and add it later. Then we take the limit $b\gg1$ and we obtain
\beq
{1\/b^2}\int^{\infty}_0dy\int^1_0dt\;{y\cos(byt)\/{e^y-1}}\mathrm{T}''_{\sigma}(t)\approx{\pi \mathrm{T}''_{\sigma}(0)\/{2b^3}}
+{1\/{2b^4}}\int^1_0dt\;{{\mathrm{T}''_{\sigma}(t)-\mathrm{T}''_{\sigma}(0)-\mathrm{T}'''_{\sigma}(0)t}\/t^2}\;.
\label{I_2-part 4}
\eeq
Computing the other integrations in Eq.~(\ref{I_2-3}) and combining the results with Eq.~(\ref{I_2-part 4}), we get the approximate result for $\mathrm{\tilde{I}}_{2\sigma}$ as
\beq
\mathrm{\tilde{I}}_{2\sigma}\approx{2c^4\/{\beta^4\omega_0}}\biggl[-{\pi^2\/6}{\mathrm{T}'_{\sigma}(0)\/b^2}-{\pi\/2}{\mathrm{T}''_{\sigma}(0)\/b^3}\biggr]
\label{final I_2 in low temp and long distance}
\eeq
up to the order $b^{-3}$ in the limit $b\gg1$.

Notice that when adding $\mathrm{\tilde{I}}_{1\sigma}$ (see Eq.~(\ref{final I_1 in low temp and long distance})) and $\mathrm{\tilde{I}}_{2\sigma}$ (see Eq.~(\ref{final I_2 in low temp and long distance})), by using the relation Eq.~(\ref{T'-A' relation}), the terms proportional to $z^{-2}$  are canceled out completely and the leading term is proportional to $z^{-3}$. This is exactly what happens when calculating Eq.~(\ref{eq for real dielectric-long distance}).

\section{ Integrals in Equations.~(\ref{E eq for real dielectric}) and (\ref{E neq for real dielectric}) in the high temperature limit }
\label{app:D}
In the high temperature limit, ${\beta\/\lambda_0}\ll1$.
The integrals  in Eqs.~(\ref{E eq for real dielectric}) and (\ref{E neq for real dielectric}) are of the same forms as those in Eqs.~(\ref{real dielectric I_1}) and (\ref{real dielectric I_2}), which can be changed to
\bea
\mathrm{\tilde{I}}_{1\sigma}&=&\biggl({c\/\beta}\biggr)^3\int^1_0dt\;\mathrm{A}_\sigma(t)
\int^{\infty}_0dy\biggl({y^3\/{y+y_{0}}}-{y^3\/{y-y_{0}}}\biggr){e^{-a y t}\/{e^y-1}}\;,
\label{real dielectric I_1-2}\\
\mathrm{\tilde{I}}_{2\sigma}&=&\biggl({c\/\beta}\biggr)^3\int^1_0dt\;\mathrm{T}_\sigma(t)
\int^{\infty}_0dy\biggl({y^3\/{y+y_{0}}}-{y^3\/{y-y_{0}}}\biggr){\cos(b y t)\/{e^y-1}}
\label{real dielectric I_2-2}
\eea
where the parameters $a$ and $b$ are the same as those defined in Appendix~{\ref{app:C} and $y_0={\beta\/\lambda_0}$.

(1)\;The asymptotic result of $\mathrm{\tilde{I}}_{1\sigma}$.

When $a\ll1$ and $ay_0\ll1$, i.e., $2z\sqrt{\epsilon-1}\ll\beta\ll\lambda_0$,
\bea
\mathrm{\tilde{I}}_{1\sigma}&=&{c^3\/{a^3\beta^3}}\int^1_0dt\;\mathrm{A}_\sigma(t)
\int^{\infty}_0dx\biggl({x^3\/{x+x_{0}}}-{x^3\/{x-x_{0}}}\biggr){e^{-x t}\/{e^{x/a}-1}}\nn\\
&\approx&{{2x_0c^3}\/{a^3\beta^3}}\int^1_0dt\;\mathrm{A}_\sigma(t)\biggl(t\int^{\infty}_0dx\;{{x^2}\/{e^{x/a}-1}}-{{t^2}\/{2}}\int^{\infty}_0dx\;{{x^3}\/{e^{x/a}-1}}\biggr)\nn\\
&\approx&4\zeta[3]\cdot {ay_0c^3\/\beta^3}\int^1_0dt\;t\mathrm{A}_\sigma(t)-{{\pi^4a^2c^2\omega_0}\/{15\beta^2}}\int^1_0dt\;t^2\mathrm{A}_\sigma(t)
\eea
in which $x_0=ay_0$ and we have only kept the $z$-dependent term.

When $a\gg1$ and $ay_0\ll1$, i.e., $\beta\ll2z\sqrt{\epsilon-1}\ll\lambda_0$, for the $y$-integration  in Eq.~(\ref{real dielectric I_1-2}), we can expand the factor $(e^{y}-1)^{-1}$ to be an infinite sum of a series, and then by changing variables, it can be re-expressed as
\bea
\mathrm{\tilde{I}}_1'&=&\int^{\infty}_0dy\biggl({y^3\/{y+y_{0}}}-{y^3\/{y-y_{0}}}\biggr){e^{-a y t/c}\/{e^y-1}}\nn\\
&=&\sum_{n=1}^{\infty}e^{ny_0}\int_{y_0}^{\infty}dy{{(y-y_0)^3}\/{y}}e^{-a(y-y_0)t}e^{-ny}\nn\\
&-&\sum_{n=1}^{\infty}e^{-ny_0}\int_{-y_0}^{\infty}dy{{(y+y_0)^3}\/{y}}e^{-a(y+y_0)t}e^{-ny}\;.\label{expand}
\eea
As $y_0\ll1$, we approximate the infinite sum in the above equation by integration.  After some simplifications, $\mathrm{\tilde{I}}'_1$ can be changed to
\beq
\mathrm{\tilde{I}}'_1=y_0^2\int_{0}^{\infty}dy\biggl({{y^2}\/{y+1}}-{{y^2}\/{y-1}}\biggr)e^{-(at+1)y_0y}\;,
\eeq
thus
\beq
\mathrm{\tilde{I}}_{1\sigma}={c\omega_0^2\/{\beta}}\int_0^1dt\;\mathrm{A}_\sigma(t)\int_{0}^{\infty}dy\biggl({{y^2}\/{y+1}}
-{{y^2}\/{y-1}}\biggr)e^{-(at+1)y_0y}\;.\label{transformed tilde I1sigma}
\eeq
Performing the $y$-integration in Eq.~(\ref{transformed tilde I1sigma}) directly and then taking the limit $a\gg1$ and $ay_0\ll1$, we get the asymptotic result
\beq
\mathrm{\tilde{I}}_{1\sigma}\approx-{2c^2\omega_0\/{a\beta^2}}\int^1_0dt\;{\mathrm{A}_{\sigma}(t)\/t}\;.
\eeq

When $a\gg1$ and $ay_0\gg1$, i.e., $2z\sqrt{\epsilon-1}\gg\lambda_0\gg\beta$, we can firstly change Eq.~(\ref{real dielectric I_1-2}) into Eq.~(\ref{transformed tilde I1sigma}), then do the $y$-integration by parts, subtract $\mathrm{A}'_{\sigma}(0)$ from the $t$-integration and later add it.  Finally,  taking the limit $ay_0\gg1$,  we obtain
\beq
\mathrm{\tilde{I}}_{1\sigma}\approx{4c^4\/{\beta^4\omega_0 a^3}}\biggl(\int_0^1dt\;{{\mathrm{A}_\sigma(t)-\mathrm{A}'_{\sigma}(0)t}\/t^3}-\mathrm{A}'_{\sigma}(0)\biggr)\;.
\eeq

(2)\;The asymptotic result of $\mathrm{\tilde{I}}_{2\sigma}$.

When $b\ll1$ and $by_0\ll1$, i.e., $2z\ll\beta\ll\lambda_0$, taking  steps as those we did in simplifying $\mathrm{\tilde{I}}_{1\sigma}$, we get
\bea
\mathrm{\tilde{I}}_{2\sigma}&\approx&{\pi^4c^3\/{15\beta^3}}b^2y_0\int^1_0dt\;t^2\mathrm{T}_\sigma(t)
\eea
where we have  kept only the leading $z$-dependent term.

When $b\gg1$ and $by_0\ll1$, i.e., $\beta\ll2z\ll\lambda_0$, we firstly change $\mathrm{\tilde{I}}_{2\sigma}$ to
\beq
\mathrm{\tilde{I}}_{2\sigma}={c\omega_0^2\/{\beta}}\int_0^1dt\;\mathrm{T}_\sigma(t)\int_{0}^{\infty}dy\biggl({{y^2}\/{y+1}}
-{{y^2}\/{y-1}}\biggr)e^{-y_0y}\cos(by_0yt)\label{transformed tilde I2sigma}
\eeq
as we have done for Eq.~(\ref{real dielectric I_1-2}) (see Eqs.~(\ref{expand})-(\ref{transformed tilde I1sigma})).  Then we divide the above double-integral into the sum of two parts as
\bea
\mathrm{\tilde{I}}_{2\sigma}&=&{c\omega_0^2\/{\beta}}\biggl[-2\int^1_0dt\;\mathrm{T}_{\sigma}(t)\int^{\infty}_0dy\;e^{-y_0y}\cos(by_0yt)\nn\\
&&\quad\;\quad\;\;+\int^1_0dt\;\mathrm{T}_{\sigma}(t)\int^{\infty}_0dy\biggl({1\/{y+1}}-{1\/{y-1}}\biggr)e^{-y_0y}\cos(by_0yt)\biggr]\;,
\label{I_2 two parts}
\eea
do the two integrals on the right hand side of the above equation directly,  and finally  take the limits $y_0\ll1$ and $by_0\ll1$. As a result, we get the asymptotic result
\beq
\mathrm{\tilde{I}}_{2\sigma}\approx-{\pi c^2\omega_0\/{\beta^2b}}\mathrm{T}_{\sigma}(0)\;.
\eeq

When $b\gg1$ and $by_0\gg1$, i.e., $2z\gg\beta\gg\lambda_0$, we firstly change $\mathrm{\tilde{I}}_{2\sigma}$ into the sum of two parts as in Eq.~(\ref{I_2 two parts}). For the first double integral on the right hand side of Eq.~(\ref{I_2 two parts}), the $y$-integration can be done directly, so only the $t$-integration is left.  For the $t$-integration, it diverges at the point $t=0$ if we take the limit $by_0\gg1$ directly.  However, we can subtract $\mathrm{T}_{\sigma}(0)$ and $\mathrm{T}'_{\sigma}(0)t$ from the $t$-integration and later add them. Similarly, for the second double integral on the right  hand side of Eq.~(\ref{I_2 two parts}), because it diverges at the point $t=0$ if we do the $y$-integration and take the limit $by_0\gg1$ directly, we can subtract $\mathrm{T}_{\sigma}(0)$ and $\mathrm{T}'_{\sigma}(0)t$ from the $t$-integration and later add them. After these steps  and further taking the limits $b\gg1$ and $by_0\gg1$, we obtain
\bea
\mathrm{\tilde{I}}_{2\sigma}&\approx&-{\pi c^2\omega_0\/{b\beta^2}}\mathrm{T}_{\sigma}(0)+{\pi c\omega_0^2\/{\beta}}\int^1_0dt\;\mathrm{T}_{\sigma}(t)\sin(by_0t)\nn\\
&\approx&-{\pi c^2\omega_0\/{b\beta^2}}\mathrm{T}_{\sigma}(1)\cos(by_0)\;.
\eea

\end{document}